
\magnification=\magstep1
\raggedbottom


\font\elevenbf=cmbx10 scaled\magstep 2
\font\lrm     = cmr10                     
\font\srm     = cmr9                      
\font\sbf     = cmbx9                     

\lrm

\def\ref{\noindent \hangafter=1 \hangindent=1.0\parindent}
\def\etal{{et al.~}}
\def\lapproxeq{\lower 0.5ex\hbox{$\; \buildrel < \over \sim \;$}
               \allowbreak}
\def\gapproxeq{\lower 0.5ex\hbox{$\; \buildrel > \over \sim \;$}
               \allowbreak}

\def\e{1E1740.7-2942}
\def\es{1E1740.7-2942~}
\def\lsim{\lower.5ex\hbox{$\; \buildrel < \over \sim $}}
\def\gsim{\lower.5ex\hbox{$\; \buildrel > \over \sim $}}

\null
\vskip 0.2 truecm

{}~~~~~~{\elevenbf Astrophysical Gamma Ray Emission Lines}\footnote{$^1$}
{\noindent In press, 1995, in {\it The Analysis of Emission
Lines}, eds. R. E. Williams and M. Livio, (Cambridge: Cambridge
Univ. Press), based on an invited talk at the {\it Analysis of
Emission Lines} conference, Space Telescope Science Institute,
May 1994.}

\vskip 0.6 truecm

\centerline{REUVEN RAMATY}

\centerline{Laboratory for High Energy Astrophysics,
   Goddard Space Flight Center}

\centerline{Greenbelt, MD 20771, USA}

\vskip 0.2 truecm
\centerline{and}
\vskip 0.2 truecm

\centerline{RICHARD E. LINGENFELTER}
\centerline{Center for Astrophysics and Space Sciences,
University of California San Diego}
\centerline{La Jolla, CA 92093, USA}

\vskip 1. truecm

\noindent {\bf ABSTRACT}
\vskip 0.2 truecm

We review the wide range of astrophysical observations of gamma
ray emission lines and we discuss their implications. We
consider line emission from solar flares, the Orion molecular
cloud complex, supernovae 1987A and 1991T, the supernova
remnants Cas A and Vela, the interstellar medium, the Galactic
center region and several Galactic black hole candidates. The
observations have important, and often unique, implications on
particle acceleration, star formation, processes of
nucleosynthesis, Galactic evolution and compact object
physics.

\vskip 0.5 truecm
\noindent {\bf 1. INTRODUCTION}
\vskip 0.2 truecm

Gamma ray lines are the signatures of nuclear and other high
energy processes occurring in a wide variety of astrophysical
sites, ranging from solar flares and the interstellar medium to
accreting black holes and supernova explosions. Their
measurement and study provide direct, and often unique,
information on many important problems in astrophysics,
including particle acceleration, star formation,
nucleosynthesis and the physics of compact objects.

The physical processes that produce astrophysical gamma ray
emission lines are nuclear deexcitation, positron annihilation
and neutron capture. Excited nuclear levels can be populated by
the decay of long-lived radioactive nuclei as well as directly
in interactions of accelerated particles with ambient gas.
Nuclear deexcitation lines following radioactive decay have
been seen from supernova 1987A (Matz et al. 1988; Tueller et
al. 1990; Kurfess et al. 1992), from the supernova remnants Cas
A (Iyudin et al. 1994) and Vela (Diehl et al. 1995), and the
interstellar medium (Mahoney et al. 1984; Share et al. 1985;
Diehl et al. 1994; 1995). The observation of such line emission
provides unique information on processes of nucleosynthesis.
Nuclear deexcitation lines following accelerated particle
interactions have been observed from solar flares (Chupp et al.
1973; Rieger 1989; Chupp 1990) and recently from the Orion
molecular cloud complex with the COMPTEL imaging spectrometer
on the Compton Gamma Ray Observatory (CGRO, Bloemen et al.
1994). The observed Orion lines, at 4.44 MeV from $^{12}$C and
6.13 MeV from $^{16}$O, cannot result from processes of
nucleosynthesis since there are no significant long lived
radioactive isotopes that decay into the excited states of
these nuclei. The lines must therefore be produced
contemporaneously by accelerated particles.

The accelerated ions which produce the deexcitation lines also
produce neutrons and positrons. The neutrons can be captured by
various nuclei. Capture on hydrogen produces deuterium and
2.223 MeV line photons. This line has been extensively observed
from solar flares (e.g. Chupp 1990). Capture on $^{56}$Fe
produces $^{57}$Fe and a variety of gamma ray lines, the most
important of which are at 7.645 and 7.631 MeV corresponding to
captures into the ground state and first excited state of
$^{57}$Fe. The observability of the neutron capture lines
requires a region in which the ambient density is high enough
to allow the capture of the neutrons before they decay and the
opacity low enough to allow the escape of the line emission.
Such a site is provided by the solar photosphere. Neutron
capture lines have not yet been seen from other astrophysical
sites, except for unconfirmed lines from a transient source
(Ling et al. 1982) which were interpreted as redshifted lines
from neutron capture on both hydrogen and iron (Lingenfelter,
Higdon, and Ramaty 1978).

Positron-electron annihilation leads to the 0.511 MeV line provided
that the temperature of the annihilation site is sufficiently low;
otherwise the annihilation radiation is broadened and blueshifted,
and at temperatures approaching $m_ec^2/k$ it is eventually smeared
into a continuum. The annihilation radiation can also be redshifted,
leading to line emission below 0.511 MeV, if the annihilation site is
sufficiently close to a neutron star or black hole. A narrow line at
precisely 0.511 MeV (with line centroid error of only a few tenths of
keV and width less than 3 keV) has been observed on many occasions
with Ge detectors from the direction of the Galactic center
(Leventhal, MacCallum, and Stang 1978; Gehrels et al. 1991; Smith et
al. 1993). This line has also been observed with the OSSE instrument
on CGRO (Purcell et al. 1993). The line centroid and width require
that the positrons annihilate in the interstellar medium, at
considerable distances from compact objects.

Line features at energies just below 0.511 MeV have been seen
from Galactic black hole candidates (e.g. Gilfanov et al. 1994;
Smith et al. 1993; Briggs et al. 1995). Their origin, however,
is not clear. They have generally been attributed to redshifted
annihilation radiation, although they could also be due to the
Compton down scattering of collimated high energy continuum
(Skibo, Dermer, and Ramaty 1994). Another line produced by
Compton scattering is that at $\sim$0.2 MeV. This line, seen
from the black hole candidate Nova Muscae (Goldwurm et al.
1992; Sunyaev et al. 1992) and  unidentified objects in the
direction of the Galactic center (Leventhal and MacCallum 1980,
Smith et al. 1993), has been identified with Compton
backscattered annihilation radiation (Lingenfelter and Hua
1991). However, the feature could also result from Compton down
scattering of higher energy continuum in a jet (Skibo et al.
1994).

Cyclotron absorption features in intense (teragauss) magnetic
fields have been observed from X-ray binaries and gamma ray
bursts. As here we deal only with emission lines, we shall not
discuss these features. Instead, we refer the reader to our
previous review (Ramaty and Lingenfelter 1994).

The plan of the present article is as follows: in section \S 2
we deal with the line emission from accelerated particle
interactions and we discuss solar flares and Orion; in \S 3 we
treat the lines from processes of nucleosynthesis and we
discuss the supernova and $^{26}$Al observations; in \S 4 we
deal with Galactic positron annihilation; in \S 5 we discuss
gamma ray line emissions from black hole candidates; and we
present our conclusions in \S 6.

\vskip 0.5 truecm \noindent {\bf 2. ACCELERATED PARTICLE
INTERACTIONS} \vskip 0.2 truecm

The interactions of accelerated particles with ambient matter
produce a variety of gamma ray lines following the deexcitation
of excited nuclei in both the ambient matter and the
accelerated particles. Deexcitation gamma ray lines can be
broad, narrow or very narrow, depending on their widths. Broad
lines are produced by accelerated C and heavier nuclei
interacting with ambient H and He. The broadening of these
lines (widths ranging from a few hundreds of keV to an MeV) is
due to the motion of the accelerated heavy particles
themselves. Narrow lines are produced by accelerated protons
and $\alpha$ particles interacting with ambient He and heavier
nuclei. The broadening in this case (widths ranging from a few
tens of keV to around 100 keV), is due to the motion of the
heavy targets which recoil with velocities much lower than
those of the projectiles. The $^7$Li and $^7$Be deexcitation
lines at 0.478 and 0.429 MeV, produced by $\alpha$ particle
interactions with ambient He, are also considered as narrow.
Very narrow lines result from excited nuclei which have slowed
down and stopped due to energy losses before emitting gamma
rays. The broadening of these lines is due only to the bulk
motion of the ambient medium (widths around a few keV or less
for the interstellar medium).

There are two distinct processes which can lead to very narrow
line emission: deexcitation of heavy nuclei embedded in dust
grains (Lingenfelter and Ramaty 1976), and deexcitation of
excited nuclei populated by long lived radionuclei. In the case
of the dust, the excitations are due to protons and $\alpha$
particles. The best example of a very narrow grain line is that
at 6.129 MeV from $^{16}$O. The mean life of the corresponding
nuclear level, 1.2 $\times$ 10$^{-11}$ s, is long enough to
allow the excited nucleus to stop in grain material before the
gamma ray is emitted. In contrast, the 4.438 MeV line of
$^{12}$C cannot be very narrow because the lifetime of the
corresponding level, 2.9 $\times$ 10$^{-14}$ s, is too short to
allow the excited nucleus to stop. In addition to a relatively
long lifetime, the production of very narrow grain lines also
requires grains that are large enough. Their size distribution
and the amount of O locked up in dust will then determine the
ratio of the very narrow to narrow 6.129 MeV line fluxes. A
ratio of about 1/3 is a reasonable average.

Long lived radionuclei produced by accelerated particle bombardment
can stop in ambient gas before they decay thereby producing excited
nuclei essentially at rest. The most important such radionuclei are
$^{55}$Co($\tau_{1/2}=17.5$h), $^{52}$Mn($\tau_{1/2}=5.7$d),
$^{7}$Be($\tau_{1/2}=53.3$d), $^{56}$Co($\tau_{1/2}=78.8$d),
$^{54}$Mn($\tau_{1/2}=312$d), $^{22}$Na($\tau_{1/2}=2.6$y), and
$^{26}$Al($\tau_{1/2}=0.72$my), all of which can be produced in
accelerated particle interactions, for example
$^{56}$Fe(p,n)$^{56}$Co. Unlike the very narrow grain lines which are
produced almost exclusively by accelerated protons and $\alpha$
particles, very narrow lines from long lived radioactivity can result
from both these interactions and interactions due to accelerated
heavy nuclei. When the interactions are predominantly due to heavy
ion interactions (as might be the case for Orion, see below), the
only narrow features in the spectrum are those from the long lived
radioisotopes. To produce a very narrow 0.847 MeV line from $^{56}$Co
it is necessary to stop a $\sim$10 MeV/nucleon $^{56}$Co in less than
about 100 days, and this requires that the density of the ambient
medium exceed about 2 $\times$ 10$^4$ cm$^{-3}$. Such densities may
be present in dense molecular clouds. Clearly the discovery of very
narrow lines would provide unique information on the density of the
medium in which the lines are formed.

A theoretical spectrum, (Fig.~1 from Ramaty 1995) illustrates some of
the above line features. The calculated spectra were obtained by
using a nuclear deexcitation line code which employs a large number
of nuclear reaction cross sections and allows calculations to be
performed for a variety of compositions, accelerated particle spectra
and interaction models (Ramaty, Kozlovsky, and Lingenfelter 1979;
Murphy et al. 1991). The top panel of Fig.~1 shows a deexcitation
line spectrum obtained by assuming a solar photospheric composition
for the ambient medium and cosmic ray source composition for the
accelerated particles. Both broad and narrow lines, as well as very
narrow lines from long lived radionuclei, are present. (Very narrow
lines from dust, however, are not included). The $^{12}$C complex
around 4.44 MeV clearly shows the narrow line superimposed on its
broad counterpart. The bottom panel of the figure shows the spectrum
obtained by suppressing the accelerated protons and $\alpha$
particles. The narrow lines are obviously absent in this case.
However, we now can clearly see the very narrow lines from the long
lived radionuclei. To allow the shortest lived radionucleus to stop
before it decays, we assumed that the ambient density exceeds $2
\times 10^6$ cm$^{-3}$.

As mentioned in the Introduction, gamma ray lines from
accelerated particle interactions were seen from the flaring
Sun and the Orion Complex. We first consider the flares.

\vskip 0.5 truecm \noindent {\bf 2.1 Solar Flares} \vskip 0.2
truecm

Gamma ray lines from solar flares were first observed in 1972
with the NaI scintillator on OSO-7 (Chupp et al. 1973). But it
was not until 1980 that routine observations of gamma ray lines
and continuum became possible with the much more sensitive NaI
spectrometer on the Solar Maximum Mission (SMM, Rieger 1989;
Chupp 1990). The SMM detector operated successfully until 1989,
making important observations during both the declining portion
of solar cycle 21 (1980-1984) and the rising portion of cycle
22 (1988-1989). Additional gamma ray line observations during
cycle 21 were carried out with a CsI spectrometer on HINOTORI
(Yoshimori 1990). During the peak of solar cycle 22 solar flare
gamma ray line observations were carried out with the CGRO
instruments OSSE (Murphy et al. 1993), COMPTEL (Ryan et al.
1993), and EGRET (Schneid et al. 1994, see also Ramaty et al.
1994), with the Phebus instrument on GRANAT (Barat et al.
1994), and with a gamma ray spectrometer on YOHKOH (Yoshimori
et al. 1994).

A theoretical solar flare gamma ray spectrum is shown in Fig.~2,
calculated for interactions of flare accelerated ions and electrons
having power law energy spectra with an index of -3.5; the calculated
spectra are normalized to the observed 4--7 MeV gamma ray flux in the
flare of 27 April 1981. These are the same parameters as those used
in our previous review (Ramaty and Lingenfelter 1994), except that
here we have limited the range of photon energies to 0.1 -- 10 MeV.
The strongest line, at 2.223 MeV from neutron capture, is discussed
separately below. The narrow deexcitation lines at 6.129 MeV from
$^{16}$O, 4.438 MeV from $^{12}$C, 1.779 MeV from $^{28}$Si, 1.634
MeV from $^{20}$Ne, 1.369 MeV from $^{24}$Mg, and 0.847 MeV from
$^{56}$Fe are clearly seen. The corresponding broad deexcitation
lines, together with a variety of other unresolved lines, form the
excess above the bremsstrahlung continuum represented by the dashed
curve. The line at 0.511 MeV is from positron annihilation, and the
excess continuum just below 0.511 MeV is due to positronium
annihilation. The strength of this continuum relative to the 0.511
MeV line depends primarily on the density of the ambient gas. These
calculations are for a positronium fraction of 0.9, which requires
that the positrons annihilate in a region of density lower than about
10$^{15}$ cm$^{-3}$ (Crannell et al. 1976). Since we assumed an
isotropic distribution of interacting particles the $^7$Li and $^7$Be
deexcitation lines at 0.478 MeV and 0.429 MeV (Kozlovsky and Ramaty
1974) blend into a single feature which peaks at $\sim$0.45 MeV, as
can be seen in the insert in Fig.~2. However, under certain
conditions of anisotropy, this feature breaks up into two individual
lines (Kozlovsky and Ramaty 1977).

As already mentioned, the 2.223 MeV line is formed by neutron
capture on $^1$H in the photosphere, at a much larger depth
than that at which the nuclear reactions take place.
Consequently, the ratio of the 2.223 MeV line fluence to the
fluence in the deexcitation lines depends on the position of
the flare on the solar disk (Wang and Ramaty 1974; Hua and
Lingenfelter 1987). The ratio becomes quite small for flares at
or near the limb, and can vanish for flares behind the limb.
This was demonstrated most dramatically by gamma ray
observations of a flare on 1 June 1991 located at 10$^\circ$
degrees behind the limb for which the 2.223 MeV line was absent
while the deexcitation lines were still seen (Barat et al.
1994). Evidently, a considerable fraction of the nuclear
reaction occurred in the corona, at a site which was visible
even though the location of the optical flare was occulted.
While this `classical' behavior is reassuring, (that the 2.223
MeV should be formed in the photosphere and hence strongly
attenuated from flares located at or behind the limb was
predicted theoretically, Wang and Ramaty 1974), there is another
observation of a flare about $10\pm5$ degrees behind the limb
for which only the 2.223 MeV line was seen (Vestrand and Forrest
1993). This flare, on 29 September 1989, was one of the largest
on record, having produced a multitude of emissions including
protons up to 25 GeV (Swinson and Shea 1990). Because of the
very strong expected attenuation, the observed 2.223 MeV line
must have been produced by charged particles interacting on the
visible hemisphere of the Sun. It was suggested (Cliver, Kahler
and Vestrand 1993) that these particles were accelerated by a
coronal shock over a large volume thereby producing an extended
gamma ray emitting region visible from the Earth even if the
optical flare was behind the limb.

The ratio of the fluence of the bremsstrahlung continuum to that in
the lines was used to determine the electron-to-proton ratio $(e/p)$
for the accelerated particles which interact at the Sun (Ramaty et
al. 1993). The derived values of $e/p$ were found to be generally
larger than the corresponding $e/p$'s obtained from observations of
solar flare particle events in interplanetary space. Such
interplanetary $e/p$ observations have been used to distinguish two
classes of solar flare particle events (Cane, McGuire and von
Rosenvinge 1986; Reames 1990). Impulsive events, for which the
associated soft X-ray emission is of relatively short duration, have
large $e/p$ ratios; gradual events, for which the soft X-rays last
longer, exhibit smaller values of $e/p$. The gamma ray results reveal
comparable or even higher values of $e/p$ than the impulsive events
in space, regardless of whether the flare is impulsive or gradual.
This result suggests that the particles which are trapped at the Sun
and produce the gamma rays, and the particles observed in
interplanetary space from impulsive flares are accelerated by the
same mechanism, probably stochastic acceleration due to gyroresonant
interactions with plasma turbulence. Relativistic electron
acceleration by such turbulence, particularly whistler waves, can be
quite efficient (Miller and Steinacker 1992). On the other hand, for
gradual events the particles are thought to be accelerated from
cooler coronal gas probably by a shock which is not expected to
accelerate electrons efficiently.

In Fig.~3 we show a solar flare gamma ray spectrum observed with the
NaI spectrometer on SMM  (Murphy et al. 1990). The $^{16}$O, $^{12}$C
and $^{20}$Ne lines can be clearly seen, but the $^{24}$Mg line is
not resolved from several neighboring features (compare with Fig.~1).
The annihilation line, possibly accompanied by a positronium
continuum, and the $\alpha$--$\alpha$ lines produce a broad feature
above the strong bremsstrahlung continuum. The 2.223 MeV line is weak
because of the location of this flare at the limb of the Sun.

The spectrum shown in Fig.~3 was used to determine abundances
in both the ambient solar atmosphere and the accelerated
particle population (Murphy et al. 1991). For the accelerated
particles the results indicate that the abundances of the heavy
elements, in particular Mg and Fe, are significantly enhanced
(relative to C and O) in comparison to their abundances in
either the photosphere or the corona. Similar enhancements are
observed in the abundances of the accelerated particles
observed from impulsive flares (Reames 1990). This supports the
conclusion mentioned above that the particles responsible for
gamma ray production and the particles observed in
interplanetary space from impulsive flares are accelerated by
the same mechanism. It has been shown (Miller and Vi\~nas 1993)
that stochastic acceleration by plasma turbulence produce these
enhancements (Reames, Meyer and von Rosenvinge 1994).

For the ambient gas, the Mg, Si and Fe abundances relative to C
and O are consistent with coronal abundances but enhanced in
comparison with photospheric abundances. The enhanced Mg, Si
and Fe abundances (elements with low first ionization
potential, FIP) could be understood in terms of a charge
dependent ambient gas transport process from the photosphere to
the chromosphere and corona which favors the collisionally
ionized, low FIP elements in the photosphere (Meyer 1985).
Indeed, the enhancement of the low FIP elements in the corona
is rather well established from various observations, in
particular solar flare accelerated particle observations. The
abundance of Ne determined from the gamma ray data is
problematic. Based primarily on the strong 1.634 MeV $^{20}$Ne
line (Fig.~2), the gamma ray data yielded (Murphy et al. 1991)
a Ne to O ratio that is more than a factor of 2 larger than the
coronal Ne/O which is thought to be consistent with the local
galactic Ne abundance (Meyer 1989). The photospheric Ne
abundance is not known, so in principle it is possible that the
gamma ray derived abundance would be representative of the
photosphere. This point of view, however, is challenged by
other astronomical data which would favor a photospheric Ne
abundance about equal to its coronal value (Meyer 1989). It has
been proposed that the Ne enhancement could be due to
photoionization by soft X-rays (Shemi 1991), an interpretation
which predicts that S should also be enhanced. Both the Ne and
S enhancements have been confirmed by soft X-ray observations
(Schmelz 1993), but only in one flare. Thus the issue of the
gamma ray derived Ne abundance remains unresolved, awaiting new
observations and their analysis.

Another isotope whose photospheric abundance is not well known
is $^3$He. It was first pointed out by Wang and Ramaty (1974)
that $^3$He in the photosphere could capture as much as one
half of the neutrons, so that the flux and time profile of the
2.223 MeV line, resulting from the capture of the other half on
$^1$H, would strongly depend on the photospheric $^3$He
abundance. Observations (Prince \etal 1983) of the time
dependent flux of the 2.223 MeV line were used (Hua and
Lingenfelter 1987) to derive a photospheric $^{3}$He/H $\simeq
2 \times 10^{-5}$, which is sufficiently low to be consistent
with the $^3$He abundance expected solely from primordial
nucleosynthesis without requiring much mixing in the solar
atmosphere.

Thus, we see that gamma ray emission lines from the Sun are
providing a wealth of new information on both abundances
and the acceleration of energetic particles in the solar flares.

\vskip 0.5 truecm \noindent {\bf 2.2 The Orion Complex} \vskip 0.2
truecm

The recent discovery of gamma ray emission lines from the Orion
giant molecular cloud complex has now revealed exciting new
particle acceleration processes in this nearest region of
recent star formation that have very important implications for
light element nucleosynthesis. Gamma ray line emission in the 3
to 7 MeV range was observed from the Orion complex (Fig.~4a)
with COMPTEL (Bloemen et al. 1994). The radiation shows
(Fig.~4b) emission peaks near 4.44 and 6.13 MeV, consistent
with the deexcitation of excited states in $^{12}$C and
$^{16}$O produced by accelerated particle interactions.
Moreover, the intensity of these lines is roughly two orders of
magnitude greater than that expected from irradiation by low
energy cosmic rays with energy density equal to that of the
local Galactic cosmic rays (Ramaty et al. 1979). Thus this
emission requires that the ambient matter in Orion, both gas
and dust, is undergoing bombardment by an unexpectedly intense,
locally accelerated, population of energetic particles.

\vskip 0.2truecm
\noindent {\it Implication of Gamma Ray Emission Lines}
\vskip 0.2truecm

We have explored (Ramaty, Kozlovsky and Lingenfelter 1995a,b)
the implications of these observations on the composition,
energy spectra and total power of the accelerated particles in
the Orion region. As we will discuss in detail, we have found
that the ratio of the measured flux in the 3 to 7 MeV band
compared to limits on the 1 to 3 MeV band place very strong
constraints on the composition of the accelerated particles,
requiring significant enrichment in C and O relative to the
heavier elements in order not to produce too much line emission
in the 1 to 3 MeV band. At the same time, the overall
energetics and efficiency of the gamma ray line production
require significant enrichment of C and O relative to H and He.
The lack of enhanced flux at higher ($>$100 MeV) energies also
requires that the accelerated particles have a much softer
energy spectrum than that of the Galactic cosmic rays. But the
spectrum of the accelerated particles should not be too soft in
order to maintain an acceptable overall energetic efficiency.

We have calculated the gamma ray line emission rate together
with the power deposited by the accelerated particles during
gamma ray production. In the thick target model which we
employ, the ratio of the photon production rate to the power
deposited by the accelerated particles depends only on the
composition and spectrum of the accelerated particles and on
the composition and state of ionization of the ambient medium.
In Fig.~5 (from Ramaty et al. 1995b) we show the deposited
power at Orion for a neutral ambient medium with solar system
abundances and for various accelerated particle compositions as
a function of the spectral parameter $E_0$ of the accelerated
particles. The spectrum
 $$N_i(E) = K_i E^{-1.5}e^{-E/E_0},  \eqno (1) $$
where $E$ is energy per nucleon and the $K_i$'s are
proportional to the assumed abundances, is predicted by
acceleration to nonrelativistic energies near a strong shock
(compression ratio $r=4$) with the effects of a finite
acceleration time or a finite shock size taken into account by
the exponential turnover (e.g. Ellison and Ramaty 1985). We
have carried out calculations for $E_0$ in the range 2--100
MeV/nucl; $E_0$, however, should not exceed about 30 MeV/nucl,
because otherwise the accelerated particles would contribute to
pion production and thus be in conflict with the high energy
gamma rays observed from Orion with EGRET; these data only
require a cosmic ray flux equal to that observed locally near
Earth (Digel, Hunter and Mukherjee 1995).

The curves in Fig.~5 correspond to various assumed accelerated
particle  compositions (Ramaty et al. 1995b): SS -- solar
system (Anders and Grevesse 1989), CRS -- cosmic ray source
(Mewaldt 1983), SN35 -- the ejecta of a 35 M$\odot$ supernova
(Weaver and Woosley 1993), WC -- the late phase wind of a
Wolf-Rayet star of spectral type WC (Maeder and Meynet 1987); GR
-- pick up ions resulting from the breakup of interstellar
dust. Concerning the dust, in analogy with the anomalous
component of the cosmic rays observed in interplanetary space
(e.g. Adams et al. 1991), we considered the effects of the pick
up of ions onto a magnetized high speed wind (e.g. the ejecta
of supernovae or the winds of massive stars). As the ions
acquire considerable energy during the pick up process, they
form a seed population that is much more easily accelerated
than the rest of the ambient plasma. In the solar system the
pick up ions are interstellar neutrals that penetrate into the
solar cavity where they are ionized. For Orion we proposed that
the incoming matter is essentially neutral dust that is broken
up by evaporation, sputtering or other processes. The GR
composition, therefore, has no He and Ne and is relatively poor
in H.  The SS and CRS compositions will thus produce combined
broad and narrow line spectra, while the WC and GR compositions
will lead to essentially pure broad line spectra. The SN35 will
still have a weak narrow line component. We see from In Fig.~5
that the gamma ray line production is energetically most
efficient for large values of $E_0$ and accelerated particle
compositions that are poor in protons and $\alpha$ particles
(i.e. the GR and WC compositions). Thus, just from energetic
arguments, the broad line spectra are preferred, even though the
COMPTEL gamma ray data can so far not distinguish between pure
broad line spectra and combined broad and narrow line spectra
(i.e. spectra produced by heavy projectiles as well as protons
and $\alpha$ particles, Ramaty et al. 1995a).

For the  GR and WC compositions and $E_0$ = 30 MeV/nucl, the
deposited power is 4x10$^{38}$ erg s$^{-1}$ and the ionization
rate is 0.3 $M_\odot$ yr$^{-1}$ or $\zeta$ = 10$^{-13}$
$M_5^{-1}$ (H atom)$^{-1}$ s$^{-1}$, where $M_5$ is the total
irradiated neutral H mass in units of 10$^5$ solar masses.
Setting the ionization rate equal to the recombination rate,
$$ \zeta n_{\rm H} =  \alpha_r n_{\rm e} n_{\rm H}, \eqno(2) $$
where $\alpha_r \simeq 10^{-11} (T/100{\rm K})^{-0.7}$ (Bates and
Dalgarno 1962), we obtain an equilibrium electron density
$n_{\rm e}$ = 0.01 $M_5^{-1}$ $(T/100{\rm K})^{0.7}$. We thus
see that, even though the rate of ionization is quite high, for
a large enough irradiated mass and temperature typical of dense
molecular clouds, the irradiated cloud can remain essentially
neutral. The total deposited power depends of course on the
duration of the irradiation. For example, if the accelerated
particle bombardment lasts $\sim$10$^{5}$ years, the total
energy requirement would be 1.2x10$^{51}$ ergs, equal to the
total mechanical output a supernova.

Just such a supernova, occurring $\sim$ 80,000 years ago in the
 OB association at $l$ = 208$^\circ$ and $b$ =
$-$18$^\circ$, the same location as the center of the gamma ray
line source, was suggested by Burrows \etal (1991) from
analyses of the X-ray emission from the Orion-Eridanus bubble.

In addition to being energetically efficient, the WC and GR
compositions have the advantage of predicting gamma ray spectra
which are consistent with the upper limit set by the COMPTEL
observations on the emission in the 1--3 MeV range. We have
shown (Ramaty et al. 1995b) that the discrepancy between this
limit and the CRS prediction is greater than $3\sigma$, and
that the predictions of the SS and SN35 compositions are also
inconsistent at greater than $2\sigma$. On the other hand, both
the GR and WC compositions yielded 1--3 MeV fluxes which are
lower than the COMPTEL upper limit. However, while the WC
composition predicts practically no emission in the 1--3 MeV
region, the GR composition predicts significant broad line
emission in this region, due to Mg, Si and Fe. The reduction in
the overall 1--3 MeV emission for the GR case is caused by the
absence of 1.634 MeV line due to the lack of Ne in grains.

The nature of the acceleration mechanism in Orion is
still very poorly understood. Acceleration by the shocks
associated with the winds of young O and B stars was proposed
by Nath and Biermann (1994), while Bykov and Bloemen (1994)
proposed that the acceleration is due to the shocks produced by
colliding stellar winds and supernova explosions. We have
proposed (Ramaty et al. 1995b) that the pick up ions resulting
from the breakup of interstellar grains could be the dominant
injection process to any of these acceleration mechanisms. We
have also emphasized the comparison with solar flares (Ramaty
1995; Ramaty et al. 1995a). The solar flare gamma ray spectra
show much higher ratios of 1--3 MeV to 3--7 MeV fluxes than
does Orion. It was shown (Murphy et al. 1991) that this
enhanced emission below 3 MeV is, in part, due to the
enrichment of the flare accelerated particle population in
heavy nuclei. Such enrichments are routinely seen in direct
observations of solar energetic particles from impulsive flares
(e.g. Reames, Meyer and von Rosenvinge 1994). These impulsive
flare  events are also rich in relativistic electrons.  On the
other hand, in gradual events the composition is coronal and
the electron-to-proton ratio is low. As we have pointed out
above, the acceleration in impulsive events is thought to be
due to gyroresonant interactions with plasma turbulence while
in gradual events it is the result of shock acceleration. The
fact that the ratio of bremsstrahlung-to-nuclear line emission
in Orion is very low lends support to the shock acceleration
scenario. Moreover, as we noted above, these shocks may be
powered by an $\sim$ 80,000 year old supernova of a massive
star in the OB association which is also responsible for
the Orion-Eridanus bubble.

\vskip 0.2truecm
\noindent {\it The Relationship to Light Isotope Production}
\vskip 0.2truecm

That cosmic ray spallation is important to the origin of the
light isotopes $^6$Li, $^7$Li, $^9$Be, $^{10}$B, and $^{11}$B
has been known for over two decades (Reeves, Fowler and Hoyle
1970). It was shown that cosmic rays with flux equal to the
observed flux near Earth, interacting with interstellar matter
prior to the formation of the solar system, can produce the
observed solar system abundances of $^6$Li, $^9$Be and $^{10}$B
(Meneguzzi, Audouze and Reeves 1971; Mitler 1972). Since low
energy cosmic rays with spectra similar to that of the
accelerated particles in Orion also produce these isotopes, the
Galactic inventories of $^6$Li, $^9$Be and $^{10}$B can set
limits on the total Galactic irradiation by such low energy
cosmic rays. In Fig.~6 (from Ramaty et al. 1995b) we show
the ratio of $^9$Be production to 3--7 MeV deexcitation photon
production. As both the 3--7 MeV photons and the $^9$Be are
produced predominantly in interactions involving C and O, this
ratio is practically independent of composition. Adopting a
total Galactic $^9$Be inventory of 10$^{57}$ atoms and assuming
that currently there are $N_{orr}$ regions in the Galaxy with
the same level of low energy cosmic ray activity as Orion (i.e.
producing 3--7 MeV nuclear gamma rays at a rate
$Q_{orr}(3-7{\rm MeV})$ = 2.3x10$^{39}$ s$^{-1}$), we have that
$$N_{orr} Q_{orr}(3-7{\rm MeV}) \bigl[Q(^9Be)/Q(3-7){\rm
MeV}\bigr] T_{irr} <
10^{57} {\rm atoms}, \eqno (3) $$
where the quantity in square brackets is from Fig.~6 and
$T_{irr}$ is the total duration of the irradiation. Taking
$Q(^9Be)/Q(3-7{\rm MeV})=8 \times 10^{-3}$ (from Fig.~6) and
$T_{irr}=3\times 10^{17}$ s, the Galactic age, we get that
$N_{orr} < 200$. Eq.~(3) assumes the irradiation is constant in
time and ignores the destruction of $^9$Be in stars. Using a
similar argument, but employing the Galactic inventory of B,
Reeves and Prantzos (1995) obtained $N_{orr} < 100$. They
obtained a lower value because the B inventory of 10$^{58}$
atoms that they used corresponds (for the solar system
B/Be) to a $^9$Be inventory which is lower by a factor of 3
than the value we used.

The upper limit on the number of currently active `Orion-like'
regions can be used to set limits on the 3--7 MeV nuclear line
emission from the central regions of the Galaxy. For a given
type of emission (e.g. the 3--7MeV nuclear line emission), the
relationship between the flux from the solid angle defined by
the central radian of Galactic longitudes and all latitudes,
$\Phi_{\rm crad}$, and the total Galactic photon luminosity,
$Q_G$, can be written as
$$\Phi_{\rm crad}(3-7{\rm MeV}) = \xi 10^{-46} Q_G(3-7{\rm MeV})
   = \xi 10^{-46} N_{orr} Q_{orr}(3-7{\rm MeV}), \eqno(4) $$
where $\xi$ is obtained by integrating the photon source
distribution along all the lines of sight within the above
solid angle. For a variety of assumed Galactic source
distributions $\xi$ was found to range from about 0.6 to 1.7
(Skibo 1993). Thus, for $N_{orr} < 200$, $\Phi_{\rm
crad}$(3--7MeV) $<$ 5x10$^{-5}\xi$ photons cm$^{-2}$ s$^{-1}$,
which is lower by at least a factor of 3 than the upper limit
on $\Phi_{\rm crad}$(3--7MeV) obtained from SMM data (Harris,
Share and Messina 1995). We note, however, that the limit on
$N_{orr}$ obtained from $^9$Be would become higher if a
significant amount of $^9$Be is destroyed by incorporation into
stars. On the other hand, the limit would be lower if the rate
of irradiation in the early Galaxy was much higher than the
average. We clearly need more observations of nuclear gamma ray
lines to map out the distribution of low energy cosmic rays in
the Galaxy.

While Galactic cosmic ray spallation can account for the
$^6$Li, $^9$Be and $^{10}$B, such cosmic rays cannot account
for the abundances of $^7$Li and $^{11}$B. The $^{11}$B excess
is probably produced by spallation, either by low energy cosmic
rays (Casse, Lehoucq and Vangioni-Flam 1995) or by neutrinos in
supernovae (Woosley et al. 1990). Here we discuss some of the
issues related to Li.

$^7$Li production in standard big bang nucleosynthesis models
leads to $^7$Li/$^1$H $\sim$ 10$^{-10}$, in agreement with the
$^7$Li abundance in extremely metal deficient Pop II stars (see
Reeves 1994 and references therein). This cosmological $^7$Li,
however, is insufficient to account for the $^7$Li abundance in
Pop I stars and meteorites, where $^7$Li/$^1$H $\sim$
10$^{-9}$. The excess $^7$Li is thought to be produced in
stars, most likely AGB stars, although neutrino induced
spallation in supernovae was also considered (Woosley et al.
1990). Thus, while cosmic ray spallation leads to $^7$Li/$^6$Li
$\simeq$ 1.4 (Reeves 1994), the enhanced meteoritic
$^7$Li/$^6$Li of 12.3 is generally understood as being due to
these additional $^7$Li sources. That the enhancement cannot be
due to low energy cosmic rays with spectrum given by eq.(1) can
be seen from Fig.~7, which shows our calculations of
$^7$Li/$^6$Li as a function of $E_0$ for the 5 assumed
compositions. For $E_0$ around 30 MeV/nucl, $^7$Li/$^6$Li is
1.5 independent of composition and essentially equal to the
cosmic rays ratio.

Rather than producing the high meteoritic $^7$Li/$^6$Li, strong
localized irradiation by low energy cosmic rays could
significantly lower this ratio in selected molecular clouds
that have undergone `Orion-type' irradiation episodes (Lemoine,
Ferlet and Vidal--Madjar 1995; Reeves and Prantzos 1995). This
argument is suggested by the recent observations of Lemoine et
al. (1995) which indicate that in the direction of $\zeta$ Oph
there are two absorbing clouds (A and B) with different Li
isotopic ratios, ($^7$Li/$^6$Li)$_A$ $\simeq$8.6 and
($^7$Li/$^6$Li)$_B$ $\simeq$1.4. However, the energetic and
ionization implications of the required massive low energy
cosmic ray irradiation have not yet been examined.

\vskip 0.2truecm
\noindent {\it The Relationship to $^{26}$Al Production}
\vskip 0.2truecm

Evidence for the existence of freshly nucleosynthesized
$^{26}$Al in the interstellar medium is provided by the
observations of the 1.809 MeV line from various directions in
the Galaxy, as we discuss in detail below (see \S3.3). Evidence
for the presence of live $^{26}$Al at the time of the formation
of the solar system was obtained from the analysis of
meteoritic material (Lee, Papanastassiou and Wasserburg 1977).
Motivated by the Orion gamma ray line observations, Clayton
(1994) suggested that both the general Galactic $^{26}$Al and
the protosolar $^{26}$Al could result from low energy cosmic
ray spallation. However, the production ratio of $^{26}$Al to
3--7 MeV line emission from such spallation reactions, $<
10^{-2}$ (Ramaty et al. 1995a), combined with the upper limit
on the Galactic 3--7 MeV line flux set by the SMM data (Harris
\etal 1995), shows that low energy cosmic rays produce at best
only 1\% of the total  Galactic $^{26}$Al.

Although Clayton and Jin (1995) now recognize that it is not
possible to produce the Galactic $^{26}$Al by spallation, they
still argue that the protosolar $^{26}$Al could have resulted
from an episode of low energy cosmic ray irradiation similar to
that currently taking place in Orion. We have shown
(Ramaty et al. 1995a), however, that the COMPTEL Orion
observations, and in particular the upper limit on the 1--3 MeV
emission, imply a significantly lower $^{26}$Al production rate
than that estimated by Clayton (1994) to suggest that the
protosolar $^{26}$Al/$^{27}$Al of 5x10$^{-5}$ could have
resulted from low energy cosmic ray bombardment. That argument,
however, depended on the total irradiated mass and on the level
of protosolar low energy cosmic ray activity which could have
been different from that in Orion.

More recently by comparing the $^9$Be and $^{26}$Al yields, we
have shown (Ramaty \etal 1995b) that an independent limit can
be set on possible protosolar spallation production of
$^{26}$Al that does not depend of these parameters. In Fig.~8
we show the ratio of $^{26}$Al to $^{9}$Be production. The
solid bar is the ratio implied by the protosolar value of
$^{26}$Al/$^{27}$Al and the SS $^{27}$Al/$^{9}$Be. We see that
for $E_0>10$ MeV/nucl (lower values are energetically very
inefficient) the SS $^9$Be abundance limits the contribution of
particle bombardment to $^{26}$Al/$^{27}$Al to less than 3 to
10\% for compositions consistent with the 1--3 MeV flux limits.
Moreover, if we take into account the facts that most of the
$^9$Be had already been produced prior to the formation of the
protosolar nebula, and that some of the $^{26}$Al must have
decayed during irradiation, these limits become much lower.
This shows that it is practically impossible to produce the
protosolar $^{26}$Al by accelerated particle bombardment, quite
independent of the amount of material irradiated.

\vskip 0.5 truecm
\noindent {\bf 3. GAMMA RAY LINES FROM NUCLEOSYNTHESIS}
\vskip 0.2 truecm

Because the sites of explosive nucleosynthesis are optically
thick to gamma-rays, only the delayed gamma ray line emission
from the decay of synthesized radionuclei can be observed from
such sites that become at least partially transparent on time
scales less than the radioactive decay mean lives. Such gamma
ray lines from the decay of $^{56}$Co and other freshly
synthesized radionuclei in supernovae were predicted by
Clayton, Colgate and Fishman (1969). The most intense lines are
from $^{56}$Ni$\rightarrow^{56}$Co$\rightarrow^{56}$Fe decay,
followed by those from $^{57}$Co$\rightarrow^{57}$Fe and
$^{44}$Ti$\rightarrow^{44}$Sc $\rightarrow^{44}$Ca.  Such
radionuclei, carried outward in the expanding supernova
envelope, decay and then deexcite by gamma ray line emission.
At early times, the gamma rays interact with the material in
the supernova and are Compton scattered down to X-ray energies
which are photoelectrically absorbed and their energy is
eventually released at longer wavelengths. However, as the
supernova expands, some of the gamma rays begin to escape
without scattering. These gamma ray lines are Doppler-broadened
by the velocity spread of the radionuclei in the expanding
nebula. The gamma ray line shapes therefore reflect the
velocity distribution within the supernova, modified by the
opacity along the line of sight, and their measurement with
high resolution spectrometers can give us information on this
distribution.

Thus, observations of these gamma ray lines provide the most
direct means of testing current models of both explosive
nucleosynthesis and the dynamics of supernova ejecta. These
gamma ray lines, coming directly from the decay of radioactive
nuclei freshly produced in supernova explosions, give a
straight forward measure of the nucleosynthetic yields of the
supernovae. Being nuclear lines, their emission rates are
directly determined from their known branching ratios and
radioactive half lives; they are not subject to the
uncertainties in the estimated excitation rates that complicate
the interpretation of atomic lines. Moreover, because of
Doppler broadening and Compton attenuation, a wealth of
information about the mass-velocity distribution of the ejecta,
and the distribution of nucleosynthetic products within it, can
be obtained (Clayton 1974, Chan and Lingenfelter 1987; 1988;
1991; Gehrels, Leventhal and MacCallum 1987; Ruiz-Lapuente et
al. 1993) from the time dependent observations of the intensity
ratios and spectral shapes of the gamma ray lines.

Type Ia supernovae, which are thought to occur in accreting
white dwarfs and are optically defined by their lack of
hydrogen envelopes, are the most luminous sources of such
lines, because the bulk of the star is explosively burned to
produce nearly $1M_{\odot}$ of $^{56}$Ni, leaving a relatively
small ($<1 M_{\odot}$) overlying envelope to obscure the
emission. On the other hand, the other Type I and the Type II
supernovae, which are optically defined by the presence of
hydrogen in their envelope, are thought to occur in the core
collapse of massive giant stars and are much less luminous
sources of such lines, because the bulk of the $^{56}$Ni core,
that is formed by explosive burning, collapses to form a
neutron star and only a small amount (typically $< 0.1
M_{\odot}$) is ejected; furthermore the gamma rays from its
decay are obscured for a much longer time by the massive
(typically $> 10 M_{\odot}$) overlying envelope.

The most intense gamma ray lines (with branching ratios) are
those from $^{56}$Co$\rightarrow^{56}$Fe decay (Huo et al.
1987) at $0.8468$ MeV $(99.9\%)$, $1.2383$ MeV $(68.4\%)$,
$2.5985$ MeV $(17.4\%)$, $1.7714$ MeV $(15.5\%)$, and $1.0378$
MeV $(14.1\%)$.  The $^{56}$Co results from the decay of
$^{56}$Ni which is the parent nucleus produced in supernovae.
$^{56}$Co decays (Huo et al. 1987) with a mean life of 111.3
days, on a time scale comparable to that required for the
supernova ejecta to become transparent to the gamma rays. The
decay of $^{56}$Ni also gives important gamma ray lines,
$0.1584$ MeV $(98.8\%)$, $0.8119$ MeV $(86.0\%)$, $0.7500$ MeV
$(49.5\%)$, $0.2695$ MeV $(36.5\%)$ and $0.4804$ MeV
$(36.5\%)$, but its much shorter mean life of $8.8$ days allows
the lines to be detectable for a shorter period of time only in
the most rapidly expanding and least massive supernovae.
Another important radioisotope is $^{57}$Co, resulting from the
decay of $^{57}$Ni produced by neutron capture onto $^{56}$Ni.
$^{57}$Co decays with a longer mean life of $392$ days,
emitting two major lines (Burrows and Bhat 1986) at $0.1221$ MeV
$(85.9\%)$ and $0.1365$ MeV $(10.3\%)$, which can be used to
study the neutron flux during nucleosynthesis (Clayton 1974).

The principal lines from $^{56}$Co and $^{57}$Co have been
observed from the Type II supernova 1987A in the nearby Large
Magellanic Cloud, and these gamma ray line observations have
already provided important information on the nucleosynthetic
yield and dynamics of that Type II supernova, as we discuss
below. Similar observations of gamma ray lines from
extragalactic Type Ia supernovae by spectrometers on the CGRO
and the planned INTEGRAL (Winkler 1994) should provide critical
tests of current models of the nucleosynthetic yield and
dynamics of these supernovae. The detection with COMPTEL of the
$^{56}$Co lines from the Type Ia supernova SN1991T in NGC4527
at a distance of about $13$ Mpc has recently been reported (Dan
Morris, oral presentation, 17 Texas Symposium, 1994). The
expected fluxes of the principal lines from $^{56}$Co decay are
such that they should be detectable (Chan and Lingenfelter
1991) with INTEGRAL from roughly half of the Type Ia supernovae
in the Virgo Cluster. The calculations of the gamma ray line
profiles have also indicated how sensitive the line shapes are
to the mass-velocity distribution in the supernova ejecta.
Thus, future gamma ray line measurements will be used as tests
of supernova models and as diagnostics of supernova structure.

\vskip 0.2 truecm
\noindent {\bf 3.1 The Type II Supernova 1987A}
\vskip 0.2 truecm

The occurrence of the Type II supernova 1987A in the Large
Magellanic Cloud, at a distance of about 50 kpc from the Earth,
the brightest supernova seen in nearly 400 years, has given us
an unprecedented opportunity to directly study supernovae
through their gamma ray emission. This supernova explosion
occurred (Arnett et al. 1989) in the star Sk $-$69 202, a B3 I
supergiant with a mass of about $16 M_{\odot}$ at the time of
the explosion. This supernova has been studied over the full
spectrum from radio to gamma rays, and extensive observations
of the time dependent gamma ray line intensities and profiles
have been made with a variety of spectrometers on balloons and
spacecraft (e.g. Matz et al. 1988; Tueller et al. 1990).
Comparisons of these observations with predicted line
intensities as a function of time (Chan and Lingenfelter 1987;
Gehrels, MacCallum and Leventhal 1987; Bussard, Burrows and The
1989) have confirmed the production $0.075 M_\odot$ of
nucleosynthetic $^{56}$Co which has been independently derived
from the bolometric luminosity (Arnett et al. 1989).

The gamma ray observations have also shown that the $^{56}$Co
is extensively mixed in the ejecta. Early calculations (Chan and
Lingenfelter 1987) of the gamma ray line emission expected from
existing models (Weaver and Woosley 1980a,b) of supernovae in a
$15 M_{\odot}$ star suggested that the line emission from
$^{56}$Co decay would not be visible until nearly 600 days
after the explosion if the $^{56}$Co was confined to the
innermost layers of the supernova ejecta. The subsequent
detection (Matz et al. 1988) of the $^{56}$Co lines at 0.847
and 1.238 MeV within 200 days after the explosion showed that
substantial mixing had occurred in the ejecta (Chan and
Lingenfelter 1988; Gehrels et al. 1988), raising the $^{56}$Co
to high levels in the ejecta where the obscuration by overlying
matter was much lower. This has led to extensive modifications
of the supernova models to explore the causes and effects of
mixing on both the explosive nucleosynthesis and the ejecta
dynamics (e.g. Nomoto et al. 1988; Pinto and Woosley 1988;
Bussard et al. 1989).

The measured 0.847 and 1.238 MeV line fluxes (Tueller et
al. 1990) after the supernova explosion have been compared (see
Fig.~9) with calculations (Pinto and Woosley 1988) for the
mixed model 10HMM, yielding a reasonably good fit. This model
is for a $16 M_{\odot}$ star, having a $6 M_{\odot}$ helium
core and $10 M_{\odot}$ blue supergiant envelope, which
explodes ejecting the matter above the silicon shell and
explosively synthesizes the 0.075 $M_{\odot}$ of $^{56}$Ni
that is required to account for the bolometric light curve. In
this model the $^{56}$Ni has been mixed out through the helium
core into the envelope in order to account for the early gamma
ray line observations. However, as can be seen in Fig.~9, there
is still a significant difference between expected model and
the observed profile of the 0.847 MeV line measured 613
days after the explosion, suggesting that even more extensive
mixing and asymmetric ejection are required to account for the
observed line shape, because the line predicted for the 10HMM
model is much too narrow, and too blueshifted to be an
acceptable fit. Clearly further study is needed of the dynamics
of the mixed and asymmetric ejecta in Type II supernovae.

The 0.122 MeV line from the decay of $^{57}$Co, resulting from
neutron capture on freshly synthesized $^{56}$Ni, has been
detected with OSSE (Kurfess et al. 1992). The observed flux of
about $10^{-4}$ photons cm$^{-2}$ s$^{-1}$ suggests that the
production ratio of $^{57}$Ni/$^{56}$Ni is about 1.5 times the
solar abundance ratio of $^{57}$Fe/$^{56}$Fe. This is quite
consistent with current model calculations (e.g. Thielemann,
Hashimoto and Nomoto 1990).

\vskip 0.2 truecm \noindent {\bf 3.2 $^{44}$Ti Decay Lines From
Cas A and Other Young Supernovae} \vskip 0.2 truecm

On a longer time scale, $^{44}$Ti decays with a mean life of
anywhere between 78 years (Frekers et al. 1983) and 96 years
(Alburger and Harbottle 1990) to $^{44}$Sc, producing gamma ray
lines (Lederer and Shirley 1978) at $67.9 (100\%)$ and $78.4$
keV $(98\%)$;  $^{44}$Sc subsequently decays with a $5.7$ hr
meanlife to $^{44}$Ca, producing a line at 1.157 MeV $(100\%)$.
The relatively long lifetime of $^{44}$Ti, together with its
lower nucleosynthetic yield, make these lines too weak to
observe from extragalactic supernovae using current detectors.
However, this longer life should allow us to observe these
lines from Galactic supernovae for several hundred years after
the explosion and thus use them to discover the most recent
supernovae in our Galaxy. Historical records have allowed us to
identify only 2 or 3 nearby Galactic supernovae within the
past 300 years. However, the estimated (van den Bergh and
Tammann 1991) Galactic supernova rate of $8.4h^2$ per 100
years, gives an expected number of between 6 and 24 Galactic
supernovae within the last 300 years assuming $0.5 < h < 1$, or
$50 < H_o < 100$ km s$^{-1}$ Mpc$^{-1}$. The detection of the
$^{44}$Ti decay lines thus can enable us to discover the
locations of all of these other recent supernovae, which were
optically obscured, and from their gamma ray locations we can
then look in radio and other bands to identify their remnants.

Line emission at 1.157 MeV has recently been measured (Iyudin
\etal 1994) with COMPTEL from the youngest known Galactic
supernova remnant, Cas A (see Fig.~10). At an estimated
(Ilovaisky and Lequeux 1972) distance of 2.8 kpc and estimated
(Fesen and Becker 1991) age of $\sim$ 310 yr, the observed
$^{44}$Sc line flux of $(7.0\pm 1.7) \times 10^{-5}$ photons
cm$^{-2}$ s$^{-1}$ corresponds to an initial $^{44}$Ti mass of
about 1.4 to 3.2$ \times 10^{-4} M_\odot$, given the present
uncertainty in the $^{44}$Ti meanlife. Such a yield is quite
consistent with that expected from current models of Cas A as a
Type Ib supernova from the core-collapse of a $\sim 20 M_\odot$
Wolf Rayet star (Ensman and Woosley 1988, Shigeyama \etal
1990).

Assuming a comparable $^{44}$Ti yield in the much more frequent
Type II supernovae, COMPTEL may be able to locate several other
younger but more distant Type II supernovae, and the planned
INTEGRAL should discover the most recent dozen or two in our
Galaxy.

\vskip 0.2 truecm
\noindent {\bf 3.3 Gamma Ray Line Emission From $^{26}$Al Decay}
\vskip 0.2 truecm

On a much longer time scale, gamma ray line radiation at 1.809 MeV
results from the decay of $^{26}$Al (mean life $1.07 \times 10^6$
years) into the first excited states of $^{26}$Mg. Because of this
long lifetime and an encouraging theoretical $^{26}$Al yield in
supernovae (Schramm 1971), we proposed (Ramaty and Lingenfelter 1977)
that this line should be the nucleosynthetic line with the best
prospects for detection. The same idea was also independently
proposed (Arnett 1977). The long lifetime is important for at least
three reasons: it allows the accumulation of $^{26}$Al in the
interstellar medium, thereby ensuring steady line emission; it allows
the escape of the nucleosynthetic $^{26}$Al from its production site
before it decays, independent of the complications introduced by the
dynamics of the region in which the explosive nucleosynthesis takes
place; and it guarantees that the width of the line is going to be
very narrow. Gamma ray lines from shorter lived isotopes, such as the
0.847 MeV line of $^{56}$Co (mean life 111 days), can be observed
from a relatively close source (e.g. supernova 1987A) only for a
short period of time and they are significantly broadened due to the
rapid expansion of the supernova envelope. The 1.809 MeV line, on the
other hand, is broadened only by the rotation of the interstellar
gas, and it is therefore very narrow (full width at half maximum
about 3 keV). A narrow intrinsic line width offers considerable
advantages for detection with a high resolution Ge instrument. It was
because of these considerations that in 1977, two years before the
launch of HEAO-3, we proposed that the 1.809 MeV line should be
detectable.

The 1.809 MeV line was indeed the first nucleosynthetic gamma
ray line to be observed, showing that nucleosynthesis is an
ongoing process in the Galaxy at the present epoch. The 1.809
MeV line was first detected (Mahoney et al. 1984) from the
direction of the Galactic center with the HEAO-3 high
resolution Ge spectrometer and was subsequently confirmed by
observations (Share et al. 1985) with the NaI detector on SMM,
and balloon borne Ge instruments (e.g. MacCallum et al. 1987).
Most recently, imaging observations (Diehl et al. 1994, 1995)
of the 1.809 MeV line have been carried out with COMPTEL. As
can be seen in Fig.~11, the COMPTEL data reveal a broad, patchy
longitude distribution, that is very different from that of the
0.511 MeV line which is strongly peaked at the Galactic center
(\S 4). This result clearly demonstrates that the two line
emissions have quite different origins.

Estimates (Mahoney et al. 1984; Skibo, Ramaty and Leventhal
1992) of the total amount of $^{26}$Al that resides in the
Galaxy range from about 1.7 to 3$M_{\odot}$, depending on
assumed model for the Galactic distribution of $^{26}$Al, the
distance to the Galactic center and the exact values of the
1.809 MeV line fluxes implied by the observations. It should be
emphasized that the derivation of a photon flux from data
obtained by wide field of view detectors such as SMM and HEAO-3
does depend on the assumed longitude and latitude distribution
of the 1.809 MeV line emission.

Although the requirement of 1.7 to 3 $M_{\odot}$ of $^{26}$Al
exceeds the originally predicted (Ramaty and Lingenfelter 1977)
yield from supernovae, recent increases in the estimates of
supernova yields of $^{26}$Al and of the present rate of Type
II supernova occurrence now suggest that such supernovae could
in fact produce $^{26}$Al at close to the observed rate. In
particular, revisions (Woosley and Weaver 1986) of the reaction
rates in core collapse models have substantially increased the
calculated yield of $^{26}$Al to $\sim 6.9\times10^{-5}
M_{\odot}$ in the ejecta of a Type II supernova. More recent
studies (Woosley et al. 1990) of hitherto neglected neutrino
induced nucleosynthesis suggest a further doubling of the yield
of $^{26}$Al in such supernovae. Such a yield combined with the
higher recent estimate (van den Bergh and Tammann 1991) of the
Galactic Type II supernova rate of $6.1h^2$ per 100 years,
gives an expected Galactic nucleosynthesis rate of $(4-8) h^2
M_{\odot}$ of $^{26}$Al per $10^6$ years. Thus Type II
supernovae alone could account for all of the observed Galactic
$^{26}$Al production for any $h > 0.5$, or $H_o > 50$ km
s$^{-1}$ Mpc$^{-1}$. The calculated $^{26}$Al yields for Type I
and other supernovae are negligible by comparison (Nomoto,
Thielemann and Yokoi 1984). Other possible sources are novae
and Wolf-Rayet stars. However, recent calculations attribute no
more than $0.5 M_\odot$ of $^{26}$Al per $10^6$ years to
Galactic novae (Prantzos 1991), and $0.2-0.3 M_\odot$ per
$10^6$ years to Wolf-Rayet stars (Paulus and Forestini 1991).
These objects, therefore, appear to produce less than $1/2$ of
the $^{26}$Al required to account for the observations.
Moreover, the COMPTEL maps of the 1.809 MeV line intensity show
an enhancement at the Vela supernova remnant (Fig.~11),
supporting a supernova origin of the $^{26}$Al, and not at
$\gamma$ Vel which would have been expected if Wolf Rayet stars
were the source.

In addition to the Galactic longitude and latitude distribution
of the 1.809 MeV line emission, information on the origin of
the $^{26}$Al could also be obtained from studies of the shape
of the line. Doppler shifts of the line centroid energy of as
much as 0.5 keV are expected due to Galactic rotation,
depending on the distribution of the $^{26}$Al and the
direction of observation (Skibo and Ramaty 1991). All of these
studies will allow a much better understanding of the origin of
the radioactive Al, and hence of the chemical evolution of the
Galaxy.

\vskip 0.2 truecm \noindent {\bf 4. GALACTIC POSITRON
ANNIHILATION RADIATION} \vskip 0.2 truecm

The 0.511 MeV line is perhaps the most important line in gamma
ray astronomy. Its study began in 1970 when a line-like feature
around 0.47 MeV, assumed to be due to positron annihilation,
was observed from the direction of the Galactic center with a
balloon borne NaI detector (Johnson, Harnden, and Haymes 1972).
Various schemes were proposed to account for the redshift.
Ramaty, Borner and Cohen (1973) suggested that the observed line
was due to gravitationally redshifted annihilation radiation
produced on the surfaces of neutron stars, while Leventhal
(1973) pointed out that the convolution of a spectrum consisting
of the 0.511 MeV line and the accompanying positronium
continuum with the response function of a detector with poor
energy resolution would lead to the apparent redshift. It was
not until 1977, that the line energy was accurately determined
with a Ge spectrometer (Leventhal, MacCallum, and Stang 1978).
The observed line center energy, $510.7 \pm 0.5$ keV, clearly
established that the radiation was due to the annihilation of
positrons. In this and all subsequent detections with Ge
spectrometers (Riegler et al. 1981; Gehrels et al. 1991;
Leventhal et al. 1993; Smith et al. 1993), the line width was
found to be very narrow (full width at half maximum $<3$ keV)
and the line center energy to be at 0.511 MeV within errors
less than a keV (Fig.~12). Reviewing the possible origins for
the line emission (Ramaty and Lingenfelter 1979), we pointed
out that the most likely sources for the positrons responsible
for the observed annihilation radiation were the radionuclei
$^{56}$Co, $^{44}$Ti and $^{26}$Al resulting from various
processes of Galactic nucleosynthesis (see also Clayton 1973).

This point of view, however, was challenged by the subsequent
HEAO-3 result, that the 0.511 MeV line flux from the direction
of the Galactic center had varied on a time scale shorter than
1/2 year (Riegler et al. 1981). Even though the significance of
 this result was weakened by a different analysis of the
HEAO-3 data (Mahoney, Ling and Wheaton 1994), confirmation for
the 0.511 MeV time variability was provided by a series of
observations carried out with balloon borne Ge detectors from
1977 through 1984. Whereas strong 0.511 MeV line emission was
seen in the 1977 and 1979 flights, only upper limits were
obtained in 1981 and 1984. The statistical significance of the
implied variability was about 3$\sigma$. Since variability on
time scales of a few years or less can not be expected if the
positrons result from multiple nucleosynthetic events, we
proposed that the bulk of the positrons should be produced at
or near a single compact object of less than a light year in
diameter, most likely a black hole in the Galactic center
region (Ramaty, Leiter and Lingenfelter 1981). We subsequently
suggested (Lingenfelter and Ramaty 1982) that the annihilating
positrons could be produced by $\gamma$$\gamma$ pair production
of the higher energy ($>m_e c^2$) continuum photons which
showed an even greater intensity variation simultaneous with
the annihilation radiation. From the observed continuum flux
and the estimated (Galactic center) distance to the source, the
required rate of pair production limited the size of the source
region, less than about 10$^9$ cm. On the basis of this result,
we pointed out that the source probably was a stellar mass
black hole that did not necessarily reside at the dynamical
center of the Galaxy.

The possibility of positron production near black holes
was strengthened by the discovery of a broad line at
$\sim$0.4 MeV from the X-ray source \e, located at an angular
distance of $0.9^{\circ}$ from the Galactic center (Bouchet et
al. 1991; Sunyaev et al. 1991; see also Gilfanov et al. 1994).
This transient line emission was observed on 13--14 October
1990 with the imaging gamma ray spectrometer SIGMA. Similar
line emission from an unidentified source or sources in the
direction of the Galactic center was also observed with non
imaging Ge detectors flown on balloons in 1977 (Leventhal et
al. 1978; Leventhal and MacCallum 1980) and 1989 (Smith et al.
1993), as well as from another unidentified source 12$^\circ$
away from the Galactic center with HEAO-1 (Briggs et al. 1995).
We discuss these further in \S 5. If these 0.4 MeV line
features are due to gravitationally redshifted positron
annihilation produced near black holes, it is reasonable to
assume that a fraction of the positrons could escape and
annihilate at large distances from the hole to produce a line
at precisely 0.511 MeV (Ramaty et al. 1992). If the pair
production near the hole is time variable, the 0.511 MeV line
flux will also vary in time, although the time scale of the
latter variability would depend on the density of the medium in
which the positrons annihilate, $\sim$10$^5/n({\rm cm}^{-3})$
yrs (Guessoum, Ramaty, and Lingenfelter 1991).

Numerous observations of 0.511 MeV line emission from the
Galactic center and the Galactic plane were carried out with
OSSE on CGRO. This observatory was launched in 1991, and there
are published 0.511 MeV line observations from July 1991 to
October 1992 (Purcell et al. 1993). OSSE is a non-imaging NaI
instrument with a relatively small field of view
($11^\circ.4\times 3^\circ.8$). Most of the OSSE observations
were carried out with the detector pointing at or close to the
Galactic center. These observations suggest that the 0.511 MeV
line emission is strongly concentrated toward the Galactic
center. None of the OSSE observations show significant time
variability, but they allow 3$\sigma$ limits on weekly
variations of $\pm$60\% and daily variations of $\pm$120\%
relative to the observed flux of $(2.5\pm0.3)\times10^{-4}$
photons cm$^{-2}$ s$^{-1}$. Thus the question of the
variability of the 0.511 MeV line flux must await future
observations.

Models for the 3 dimensional Galactic distribution of
annihilation radiation based on the OSSE and other 0.511 MeV
line observations have been developed (Skibo 1993; Ramaty,
Skibo and Lingenfelter 1994). The calculated longitude
distributions (integrated over all Galactic latitudes) for two
of the models are shown in Fig.~13. The dashed curve is based
on the nova distribution of Higdon and Fowler (1989) which
consists of two morphological parts, a disk and a spheroid
centered at the Galactic center. Because novae occur on
accreting white dwarfs that eventually become Type Ia
supernovae, and because such supernovae could be important
positron sources, this distribution should provide a reasonable
starting point for the analysis. But as demonstrated by Skibo
(1993) and Ramaty, Skibo and Lingenfelter (1994), this
distribution cannot account for the very strong concentration
of the 0.511 MeV line emission at the Galactic center
(P$\sim$10$^{-6}$, Skibo 1993). The solid curve was obtained by
adding to the nova distribution an additional spheroid, which,
for simplicity was assumed to have the same shape as the
spheroid associated with the nova distribution itself. We refer
to the nova part and the additional spheroidal part of the
total distribution as the Galactic plane and the central
Galactic components, respectively. By allowing the ratio of
positron production in these two components to vary, a good fit
to the data was obtained (P$\sim$0.6, Skibo 1993) for a central
Galactic--to--Galactic plane positron production ratio of 2.6.
Normalization to the 0.511 MeV line observations also
determines the absolute productions, $2.6\times 10^{43}$ e$^+$
s$^{-1}$ and $1\times 10^{43}$ e$^+$ s$^{-1}$, for the two
components respectively.

The bulk of the positrons responsible for the Galactic plane
component could result from the decay of $^{56}$Co, $^{44}$Sc
and $^{26}$Al produced in various Galactic processes of
nucleosynthesis (e.g. Lingenfelter, Chan and Ramaty 1993). The
contribution of other processes (e.g. cosmic ray interactions,
pair production in pulsars) is quite small (Ramaty and
Lingenfelter 1991). The total Galactic positron production rate
from the decay of $^{26}$Al was estimated (Skibo et al. 1992)
to be about $0.2 \times 10^{43}$ positrons s$^{-1}$ (see also
\S 3), i.e. about 20$\%$ the total Galactic plane component.
The rest of the positrons probably result from $^{56}$Co and
$^{44}$Sc, where the total production rate of these isotopes
scales with the present rate of Galactic iron nucleosynthesis
and the relative contributions of these two isotopes depends on
the positron escape fraction from the envelopes of Type Ia
supernovae (Chan and Lingenfelter 1993). These two radionuclei
are thought to be produced primarily in Type Ia supernovae and
the distribution of their production rate is expected to follow
that of Galactic novae.

The origin of the positrons in the central component is
essentially unknown. It is possible that black hole candidates
are a major contributor to positron production in the Galactic
center region at the present epoch. Because both the positron
annihilation time and the distance that the positrons can
travel from their sources to their annihilation site is
strongly dependent on the properties of the medium in which
they are trapped, both the spatial extent of annihilation
radiation produced by a point source of positrons and the time
dependence of the emission are highly uncertain. It is possible
that the entire central Galactic component is fed by just \e.
But additional sources may also contribute significantly (see
Ramaty and Lingenfelter 1994 for more details). We might expect
comparable enhancements of the 0.511 MeV emission in the
direction of Nova Muscae and other candidate black hole
sources, although the production of a narrow 0.511 MeV line
depends on the existence of not only a positron source but also
a sufficiently dense annihilation site.  Clearly much better
mapping of the Galactic annihilation radiation by the planned
INTEGRAL and perhaps other missions is needed to resolve the
question of its origin.

\vskip 0.2 truecm \noindent {\bf 5. LINE FEATURES FROM BLACK HOLE
CANDIDATES} \vskip 0.2 truecm

As mentioned above, line-like emission features at $\sim$0.4
MeV have been observed from a number of sources assumed to be
accreting black holes, and in several of these observations
this line was accompanied by another line at $\sim$0.2 MeV. Line
emission at both $\sim$0.48 MeV and $\sim$0.19 MeV were
simultaneously observed with SIGMA from Nova Muscae during its
outburst on 20 January 1991 (Goldwurm et al. 1992). Similar
features at 0.496 MeV and $\sim$0.17 MeV and at $\sim$0.40 MeV
and $\sim$0.16 MeV were observed with balloon-borne Ge
spectrometers from an unidentified source, or sources, in the
Galactic center region in 1977 (Leventhal and MacCallum 1980)
and 1989 (Smith \etal 1993), respectively. Only the higher
energy line at $\sim$0.40 MeV was observed with SIGMA from \es
during an outburst on 13--14 October 1990 (Bouchet et al. 1991;
Gilfanov et al. 1994), and at $\sim$ 0.46 MeV with HEAO-1 from
another source, possibly the low mass X-ray binary 1H1822--371,
about 12$^\circ$ away from the Galactic center (Briggs et al.
1995). Two other transients exhibiting emission features were
observed with SIGMA from \es (Cordier et al. 1993; Gilfanov et
al. 1994). However, for one of these, that on 19-20 January
1992, the flux reported by SIGMA is in conflict at more than
3$\sigma$ with co-temporal CGRO OSSE and BATSE observations
(Jung et al. 1995; Smith et al. 1995).

The $\sim$0.4 MeV line has frequently been assumed to be
redshifted positron annihilation radiation. If the redshift is
gravitational, the line must be formed around a compact object,
presumably an accreting black hole, at distances varying from a
few Schwarzschild radii for the \es source, to more than 10
Schwarzschild radii for the 1977 source and Nova Muscae. The
$\sim$0.2 MeV line, observed at the same time from the latter
two sources, can been interpreted as Compton backscattered
reflection of the annihilation feature from the inner edge of
an optically thick accretion disk (Lingenfelter and Hua 1991;
Hua and Lingenfelter 1993).

This second line results from the fact that Compton scattering
of photons of energy $E_0$ into an angle $\theta$ will produce
photons of energy
$$E_s = {E_0 \over 1 + {E_0 \over
m_ec^2}(1-{\rm cos}\theta)},  \eqno (5)$$
so that, if the initial photons form a line at $E_0 \simeq
m_ec^2$, then backscattered photons (${\rm cos}\theta$=--1)
will form another line at $E_s \simeq m_ec^2/3 = 0.17$ MeV. The
intensity of the backscattered line is typically only $\sim$
5\% of that of the initial line, while the observed ratio of
these lines is about 30\% for Nova Muscae. However, the initial
line can be strongly attenuated by the accretion disk itself.
Hua and Lingenfelter (1993) have shown that such scattering
gives excellent agreement with the observed pairs of line-like
features which were seen from both Nova Muscae (Fig.~14) and
the 1977 source (Lingenfelter and Hua 1991). Moreover, assuming
that the observing angle of 68$\pm 14^\circ$ relative to the
axis of the accretion disk, determined for Compton scattering
in Nova Muscae, is the same as the inclination angle of the
binary system, a central black hole mass of 5.6$\pm$1.3
$M_\odot$ can be determined using the optically determined mass
function (Remillard, McClintock, and Bailyn 1992). However, the
temperature $kT$ of the gas which backscatters the photons
should not exceed about 10 keV since otherwise the
backscattered feature will be broader than observed.

On the other hand, for the \e, 1989 and 1H1822--371 sources,
the higher energy lines were observed at energies implying much
larger redshifts. This would imply that pair production and
annihilation occur at essentially the same physical site, which
leads to a problem because the temperature required to produce
the pairs greatly exceeds the upper limit on the temperature
set by the width of the $\sim$0.4 MeV line. This argument has
been quantified by Maciolek-Niedzwiecky and Zdziarski (1994) who showed
that the line centroid requires that the positrons annihilate
in a region around 3 Schwarzschild radii from the hole and that
the line width requires that the temperature in this region be
($kT\lsim$ 50 keV).

A different explanation of the redshift, assuming positron
annihilation, is that the observed $\sim$0.4 MeV line energy is due
to the motion of the annihilating region (Misra and Melia 1993) where
the pairs annihilate at the base of a jet emanating from the black
hole. This explanation would allow the separation of the annihilation
region from the pair production region. While this scenario could
provide an explanation for the \es observation, which revealed only
the higher energy line, it might not explain the 1989 data, for which
the lower energy line was more intense than that at higher energies;
is not clear whether the annihilation line could be greatly
attenuated relative to the backscattered line in such a jet geometry.

An alternative interpretation (Skibo et al. 1994) for both the
$\sim$0.2 and $\sim$0.4 MeV line features is that they result
form Compton scattering of high energy continuum photons in a
jet. This can again be seen from Eq.~(5). If $E_0>>m_ec^2$,
then $E_s \simeq m_e c^2/(1-cos\theta)$, i.e. energy of the
scattered photon is independent of its initial energy and
depends only on the scattering angle. Consequently, if the
original spectrum were very flat, or consisted entirely of
photons well above $m_e c^2$, the scattered photons would
accumulate at $E_s$, producing a line-like feature. Skibo et
al. (1994) showed that such Compton scattering in a double
sided jet can produce the two lines around 0.4 and 0.2 MeV for
jet bulk flow velocities $\beta$ around 0.55 and observing
angle cosines in the range 0.2 to 0.6. The required observing
angle for Nova Muscae for this model is 68$^\circ$ (J.G. Skibo,
private communication 1994), essentially the same as that for
the backscatter model, so that the implied black hole mass is
nearly identical. Also radio observations show that \es is in
fact associated with a double sided jet (Mirabel et al. 1992).

Lastly, a broad line in the 0.4--0.5 MeV can also be produced
by interactions amongst $\alpha$ particles, $^4$He($\alpha
,p$)$^7$Li$^{*478}$ and $^4$He($\alpha ,n$)$^7$He$^{*429}$
(Kozlovsky and Ramaty 1974). This feature has been seen from
solar flares (\S 2). However, the expected line centroid,
without any redshift, is at 0.45 MeV and this is in marginal
conflict with the Nova Muscae observation, for which the line
centroid was at 0.48 $\pm$ 0.02 MeV (Goldwurm et al. 1992), and
definitely with the 1977 observation for which the line
centroid was at 0.496 MeV. On the other hand, the observations
(Martin et al. 1992; 1994) of high lithium abundances in the
binary companions of the black hole candidates V404 Cygni, Cen
X-4 and A0620, possibly due to $\alpha$--$\alpha$ reactions in
the accretion disks of the holes, do provide support for this
model. Again Compton backscattering of these $\alpha$--$\alpha$
photons from the inner region of an optically thick
accretion disk could account for the $\lsim$0.2 MeV line
emission.

Fortunately, there are tests to distinguish between the various
models. The principal prediction of the
annihilation/backscattering model is that the two lines should
always be just below 0.511 MeV and around 0.2 MeV for all
sources. This seems to be borne out by the current, albeit very
limited, source population. On the contrary, the jet model in
which both line are due to Compton scattering, does predict
lines at various energies, including energies above 0.511 MeV.
The observation of a broad line around 1 MeV from Cygnus X-1
(Ling and Wheaton 1989) may be due Compton scattering of high
energy photons into viewing angles close to the forward
direction. Polarization observations could also distinguish the
models. For Compton scattering in jets, both lines should be
polarized (see Skibo et al. 1994); however, no polarization is
expected in either of these lines for the
annihilation/backscattering model. (Unpolarized radiation
undergoing Compton scattering becomes polarized, except in the
backward direction.) The required polarization observations
have not yet been carried out.

The $\alpha$--$\alpha$ interaction model makes two important
predictions. The first one concerns the isotopic ratio
$^7$Li/$^6$Li in the binary companions of the black hole
candidates. A ratio around 1.5 would favor the
$\alpha$--$\alpha$ reactions; a significantly higher ratio
would probably require another mechanism for producing the Li.
The second one involves the very narrow delayed 0.478 MeV line
from $^7$Be decay. For an outburst similar to that of Nova
Muscae on 20 January 1991, we expect a delayed flux of $\sim$
4$\times$10$^{-6}$ photons cm$^{-2}$ s$^{-1}$, equal to the
ratio of $^7$Be to $^7$Li production of $\sim$ 1 (Ramaty et al.
1979) times the 10.4\% branching ratio times the observed broad
line flux of 6$\times$10$^{-3}$ photons/cm$^2$sec  times its
0.5 day duration divided by the 77 day $^7$Be meanlife. Such a
narrow line flux could be detected by the planned INTEGRAL
spectrometer.

\vskip 0.2 truecm \noindent {\bf 6. CONCLUSIONS} \vskip 0.2 truecm

As we have seen, the study of gamma ray line radiation indeed
spans a broad range of astrophysical problems. In solar flares
the line radiation addresses problems of particle acceleration
and transport, particle trapping, magnetic field structure,
plasma turbulence and solar atmospheric composition. Of all
astrophysical sites, the richest gamma ray line spectra
observed so far are those from the Sun.

Deexcitation gamma ray line emission has been observed from the
Orion star formation region showing that the gas and dust in
this molecular cloud complex are currently undergoing strong
irradiation by low energy cosmic rays. Given that the duration
of the irradiation is about 10$^5$ years, the time span since a
possible recent supernova in Orion, the total energy in
accelerated particles is about 10$^{51}$ erg. The relationship
of such irradiation to the question of the origin of the light
elements is currently under investigation. However, neither the
overall Galactic $^{26}$Al nor the $^{26}$Al that was present
at the formation of the solar system could have originated in
accelerated particle bombardment.

Gamma ray lines from processes of nucleosynthesis have been
observed from a variety of sites. Observations of the $^{56}$Co
lines from supernova 1987A have shown that the supernova
explosion is much more complex than previously thought.
Specifically, the early appearance of the lines requires
mixing, while the shapes of the lines, determined with a high
resolution spectrometer, require asymmetric supernova ejecta.
Recently, gamma ray line emission from $^{44}$Ti decay has been
observed from Cas A, opening the possibility of observing
similar line emission from other young, and hitherto unknown,
supernova remnants in our Galaxy. The 1.809 MeV line from
$^{26}$Al decay has been observed and imaged. The bulk of the
2 M$_\odot$ of $^{26}$Al in the interstellar medium are thought
to be due to Type II supernovae which exploded in the Galaxy in
the last million years. A supernova origin is supported by the
observation of enhanced 1.809 MeV line emission from the Vela
supernova remnant.

Galactic annihilation radiation has been observed. The high
resolution Ge detectors flown on balloons have shown that there
is strong, very narrow line emission produced at precisely
0.511 MeV at or near the Galactic center. The recent OSSE
mapping of the flux in this line has shown that there is much
emission strongly concentrated within less than a few degrees
from the Galactic center. We believe that this concentration
points to a black hole origin, specifically pair production in
photon-photon collisions near the hole and escape and
annihilation of the positrons in the surrounding medium. In
addition, observations of the diffuse Galactic annihilation
radiation set important constraints on the current rate of iron
nucleosynthesis in the Galaxy.

Transient emission lines from black hole candidates have been
discovered. Such lines were seen from \es during an outburst in
1990, from Nova Muscae during an outburst in 1991, and from
several other unidentified sources. The interpretations involve
redshifted pair annihilation, Compton backscattering, Compton
downscattering of collimated high energy gamma ray continuum
and possibly $\alpha$--$\alpha$ interactions producing $^7$Li
and $^7$Be. The various models make predictions that will be
tested by future observations. Because of their low duty cycle,
these transients can be best studied by continuously monitoring
the Galactic plane with wide field of view gamma ray
spectrometers. These lines are providing unique probes of the
accretion disks and jets of black holes.

We wish to acknowledge Hans Bloemen and Roland Diehl for
supplying us with postscript figures of the COMPTEL data. We
also acknowledge financial support from NASA under Grant
NAG5-2811.


\def\xviiiicrc{{\it 18th Internat. Cosmic Ray Conf. Papers}}

\def\xxiiiicrc{{\it 23rd Internat. Cosmic Ray Conf. Papers}}
\def\apj{{\it ApJ}}
\def\aa{{\it Astr. and Ap.}}
\def\aas{{\it Astr. and Ap. Suppl.}}
\def\apjs{{\it ApJ Supp.}}

\def\ass{{\it Ap. and Sp. Sci.}}
\def\sp{{\it Solar Phys.}}

\def\araa{{\it Ann. Rev. Astr. and Ap.}}

\def\nature{{\it Nature}}

\def\asr{{\it Adv. Space Res.~}}
\def\pr{{\it Phys. Rev.}}

\def\ljaip{{\it Gamma Ray Transients and Related Astrophysical
   Phenomena,} eds. R. E. Lingenfelter, H. S. Hudson and D. M. Worrall
   (New York: AIP)}
\def\citaip{{\it The Galactic Center}, eds. G. R. Riegler and R.
   D. Blandford (New York: AIP)}

\def\washaip{{\it Nuclear Spectroscopy of Astrophysical
    Sources}, eds. N. Gehrels and G. H. Share, (New York: AIP)}
\def\minaip{{\it Cosmic Abundances of Matter}, ed. C. J.
    Waddington, (New York: AIP)}
\def\sacaip{{\it Gamma Ray Line Astrophysics}, eds. P.
    Durouchoux and N. Prantzos, (New York: AIP)}

\def\mnras{{\it Mon. Not. Roy. Astron. Soc.}}


\vskip 0.2 truecm \noindent {\bf 7. REFERENCES} \vskip 0.2
truecm

\srm

\ref Adams, J. H., et al. 1991, \apj, {\sbf 375}, L45.

\ref Alburger, D. E., and Harbottle, G. 1990, {\it Phys. Rev. C,}
{\sbf 41,} 2320.

\ref Anders, E., and Grevesse, N. 1989. {\it Geochim. et
Cosmochim. Acta,} {\sbf 53} 197.

\ref Arnett, W. D. 1977, {\it Ann. NY Acad.  Sci.}, {\sbf 302},
90.

\ref Arnett, W. D., Bahcall, J. N., Kirshner, R. P., and
Woosley, S. E. 1989, \araa, {\sbf 27}, 629.

\ref Barat, C. et al. 1994, \apj, {\sbf 425}, L109.

\ref Bates, D. R., and Dalgarno, A., 1962, in {\it Atomic and
Molecular Processes}, ed. D. R. Bates, (New York: Academic
Press).

\ref Bloemen, H., et al. 1994. \aa, {\sbf 281}, L5.

\ref Bouchet, L. et al. 1991, \apj, {\sbf 383}, L45.

\ref Briggs, M. S., Gruber, D. E., Matteson, J. L., and
Peterson, L. E. 1995, \apj, in press.


\ref Burrows, D. N., Singh, K. P., Nousek, J. A., Garmire, G.
P., and Good, J. 1993, \apj, {\sbf 406,} 97.

\ref Burrows, T. W., and Bhat, M. R., 1986, {\it Nuclear Data
Sheets}, {\sbf 47}, 1.

\ref Bussard, R. W., Burrows, A., and The, L. S. 1989, \apj,
{\sbf 341}, 401.

\ref Bykov, A., and Bloemen, H. 1994, \aa, {\sbf 283,} L1.

\ref Cane, H. V., McGuire, R. E., and von Rosenvinge, T. T. 1986,
\apj, 310, 448

\ref Casse, M., Lehoucq, R., and Vangioni-Flam, E., 1995,
\nature, {\sbf 373}, 318.

\ref Chan, K-W., and Lingenfelter, R. E. 1987, \apj, {\sbf 318},
L51.

\ref Chan, K-W., and Lingenfelter, R. E. 1988, in {\it Nucl.
Spectr. of Astrophysical Sources}, eds. N. Gehrels and G.
H.Share, 110.

\ref Chan, K-W., and Lingenfelter, R. E. 1991, \apj, {\sbf
368}, 515.

\ref Chan, K-W., and Lingenfelter, R. E. 1993, \apj, {\sbf 405}, 614.

\ref Chupp, E. L. 1990, {\it Physica Scripta,} {\sbf T18,} 15.

\ref Chupp, E. L. Forrest, D. J. Higbie, P. R. Suri, A. N. Tsai, C.,
and Dunphy, P. P. 1973, \nature, {\sbf 241}, 333.

\ref Clayton, D. D., Nature Phys. Sci., 1973, {\sbf 244}, 1973.

\ref Clayton, D. D. 1974, \apj, {\sbf 188}, 155.

\ref Clayton, D. D. 1994, \nature, {\sbf 368,} 222.

\ref Clayton, D. D., Colgate, S. A. and Fishman, G. J.
1969, \apj, {\sbf 155}, 75.

\ref Clayton, D. D., and Jin, L., 1995, preprint.

\ref Cliver, E. W, Kahler, S. W., and Vestrand, W. T.
1993, \xxiiiicrc, {\sbf 3}, 91.

\ref Cordier, B. et al. 1993, \aa, {\sbf 275}, L1.

\ref Crannell, C. J., Joyce, G., Ramaty, R., and Werntz, C.
1976, \apj, {\sbf 210}, 582.

\ref Diehl, R. \etal 1994, \apjs, {\sbf 92,} 429.

\ref Diehl, R. \etal 1995, \aa, in press.

\ref Digel, S. W., S. D. Hunter and R. Mukherjee. 1995, \apj,
{\sbf 441}, 270.

\ref Ellison, D. C., and Ramaty, R. 1985, \apj, {\sbf 298}, 400.

\ref Ensman, L. M., and Woosley, S. E. 1988, \apj, {\sbf 333,} 754.

\ref Fesen, R.A., and Becker, R.H. 1990, \apj, {\sbf 371,} 621.

\ref Frekers, D. \etal 1983, \pr, {\sbf C28}, 756.

\ref Gehrels, N., Barthelmy, S. D., Teegarden, B. J., Tueller,
J., Leventhal, M.,  and MacCallum, C. J. 1991, \apj, {\sbf 375},
L13.

\ref Gehrels, N., Leventhal, M., and MacCallum, C. J. 1987,
\apj, {\sbf 322}, 215.

\ref Gehrels, N., MacCallum, C. J., and Leventhal, M. 1987,
\apj, {\sbf 320}, L19.

\ref Gilfanov, M. et al. 1994, \apjs, {\sbf 92}, 411.

\ref Goldwurm, A. \etal 1992, \apj, {\sbf 389}, L79.


\ref Guessoum, N., Ramaty, R., and Lingenfelter, R. E. 1991, \apj, {\sbf 378,}
170.

\ref Harris, M., Share, G. H. and Messina, D. C. 1995, \apj, in press.

\ref Higdon, J. C., and Fowler, W. A. 1989, \apj, {\sbf 339}, 956.

\ref  Hua, X-M., and Lingenfelter, R. E. 1987, \apj, {\sbf 319}, 555.

\ref Hua, X-M., and Lingenfelter, R. E., 1993, \apj, {\sbf 416}, L17.

\ref Huo, J. D., Hu, D. L., Zhou, C. M., Han, X. L., Hu, B. H.,
and Wu, Y. D. 1987, {\it Nuclear Data Sheets,} {\sbf 51}, 1.

\ref Ilovaisky, S. A., and Lequeux, J. 1972, \aa, {\sbf 18}, 169.

\ref Iyudin, A. F. \etal 1994, \aa, {\sbf 284}, L1.

\ref Johnson, W. N., Harnden, and R. C. Haymes, 1972, \apj,
{\sbf 172}, L1.

\ref Jung, G. V. et al. 1995, \aa, in press.

\ref Kozlovsky, B., Lingenfelter, R. E., and Ramaty, R. 1987, \apj,
{\sbf 316,} 801.

\ref Kozlovsky, B., and Ramaty, R., 1974, \apj, {\sbf 191}, L43.

\ref Kozlovsky, B., and Ramaty, R., 1977, Astrophys. Letters, {\sbf 19}, 19.


\ref Kurfess, J. D. \etal 1992, \apj, {\sbf 399}, L137.

\ref Lederer, C. M., and Shirley, V. 1978, {\it Tables of
Isotopes}, (New York: Wiley).

\ref Lee, T., Papanastassiou, D. A., and Wasserburg, G. J., 1977,
\apj, {\sbf 211,} L107.

\ref Lemoine, M., Ferlet, R., and Vidal-Madjar, A. 1995, \aa, in press.

\ref Leventhal, M., 1973, \apj, {\sbf 183}, L147.

\ref Leventhal, M., Barthelmy, S. D., Gehrels, N., Teegarden, B. J.,
Tueller, J., and Bartlett, L. M. 1993, \apj, {\sbf 405}, L25.


\ref Leventhal, M., and MacCallum, C. J., 1980, {\it Ann. NY Acad. Sci.},
{\sbf 336}, 248.

\ref Leventhal, M., MacCallum, C. J., and Stang, P. D., 1978, \apj,
{\sbf 225}, L11.

\ref Ling, J. C., Mahoney, W. A., Willet, J. B.,  and Jacobson, A. S.
1982, in \ljaip, 143.

\ref Ling, J. C., and Wheaton, W. A. 1989, \apj, {\sbf 343}, L57.

\ref Lingenfelter, R. E., Chan, K-W.,  and Ramaty, R. 1993, {\it
Physics Reports}, {\sbf 227}, 133.

\ref Lingenfelter, R. E., Higdon, J. C., and Ramaty, R. 1978, in {\it
Gamma Ray Spectroscopy in Astrophysics,} eds. T. Cline and R. Ramaty,
(NASA), 252.

\ref Lingenfelter, R. E., and  Hua, X-M. 1991, \apj, {\sbf 381},
426.

\ref Lingenfelter, R. E. and Ramaty, R. 1976, \apj, {\sbf 211,} L19.

\ref Lingenfelter, R. E., and Ramaty, R. 1982, in \citaip, 148.

\ref MacCallum, C. J., Huters, A. F., Stang, P. D., Leventhal,
M. 1987, \apj, {\sbf 317}, 877.

\ref Maciolek-Niedzwiecky, A., and Zdziarski, A. 1995, \apj, in press.

\ref Maeder, A. and Meynet, G., 1987, \aa, {\sbf 182,} 243.

\ref Mahoney, W. A.,  Ling, J. C., and Wheaton, W. A. 1994,
\apjs, {\sbf 92}, 387.

\ref Mahoney, W. A.,  Ling, J. C.,  Wheaton, W. A.,  and
Jacobson, A. S. 1984, \apj, {\sbf 286}, 578.



\ref Martin, E. L., Rebolo, R., Casares, J., and Charles, P.,
A. 1992, \nature, {\sbf 358,} 129.

\ref Martin, E. L., Rebolo, R., Casares, J., and Charles, P.,
A. 1994, \apj, {\sbf 435}, 791.  \ref Matz, S. M. \etal 1988,
\nature, {\sbf 331}, 416.

\ref Meneguzzi, M., Audouze, J., and Reeves, H. 1971, \aa, {\sbf 15,} 337.

\ref Mewaldt, R. A. 1983, {\it Revs. Geophys. Space Phys.,} {\sbf 21,} 295.

\ref Meyer, J-P. 1985, \apjs, {\sbf 57}, 151.

\ref Meyer, J-P. 1989, \minaip, 245.

\ref Miller, J. A., and Steinacker, J. 1992, \apj,
{\sbf 399}, 284.

\ref Miller, J. A., and Vi\~nas, A. 1993, \apj, {\sbf 412}, 386.

\ref Mirabel, I. F., Rodriguez, L. F., Cordier, B., Paul, J.,
and Lebrun, F. 1992, \nature, {\sbf 358}, 215.

\ref Misra, R., and Melia, F., 1993, \apj, {\sbf 419}, L25.

\ref Mitler, H. E. 1972, \aas, {\sbf 17},
186.

\ref Murphy, R. J. et al. 1993, \xxiiiicrc, 3, 99.

\ref Murphy, R. J., Kozlovsky, B., and Ramaty, R. 1990, \apj,
{\sbf 331}, 1029.

\ref Murphy, R. J., Ramaty, R., Kozlovsky, B., and Reames, D.
V.  1991, \apj, {\sbf 371}, 793.

\ref Murphy, R. J., Share, G. H., Letaw,J. R., and Forrest, D. J.
1990, \apj, {\sbf 358}, 298.

\ref Nath, B. B. and Biermann, P. 1994, \mnras, {\sbf 270,} L33.

\ref Nomoto, K., Thielemann, F-K. and Yokoi, K. 1984, \apj,
{\sbf 286}, 644.

\ref Nomoto, K. \etal 1988, in {\it Physics of Neutron Stars
and Black Holes}, ed. Y. Tanaka (Tokyo:  Universal Academy Press),
441.

\ref Paulus, G., and Forestini, M. 1991, in \sacaip, 183.

\ref Pinto, P. A., and Woosley, S. E. 1988, \nature, {\sbf
333}, 534.

\ref Prantzos, N. 1991, in \sacaip, 129.

\ref Prince. T. A. et al. 1983, \xviiiicrc, {\sbf 4}, 79.

\ref Purcell, W. R. \etal 1993, \apj, {\sbf 413}, L85.

\ref Ramaty, R. 1995, in {\it The Gamma Ray Sky with COMPTON GRO and
SIGMA}, eds. M. Signore, P. Salati, and G. Vedrenne, (Dordrecht:
Kluwer), in press.

\ref Ramaty, R., Borner, G., and Cohen, J. M. 1973, \apj, {\sbf
181}, 891.

\ref Ramaty, R., Kozlovsky, B.,  and Lingenfelter, R. E. 1979, \apjs,
{\sbf 40}, 487.

\ref Ramaty, R., Kozlovsky, B.,  and Lingenfelter, R. E. 1995a, \apj,
{\sbf 438,} L21.

\ref Ramaty, R., Kozlovsky, B.,  and Lingenfelter, R. E. 1995b,
{\it Ann. NY Acad. Sci.}, in press.

\ref Ramaty, R., Leiter, D., and Lingenfelter, R. E. 1981,
{\it Ann. NY Acad. Sci.}, {\sbf 375}, 338.

\ref Ramaty, R., Leventhal, M., Chan, K-W., and
Lingenfelter, R. E. 1992,\apj, 392, L63.

\ref Ramaty, R., and Lingenfelter, R. E. 1977, \apj, {\sbf 213}, L5.

\ref Ramaty, R., and Lingenfelter, R. E. 1979, \nature, {\sbf 278,} 127.

\ref Ramaty, R., and Lingenfelter, R. E. 1991, in \sacaip, 67.

\ref Ramaty, R., and Lingenfelter, R. E. 1994, in {\it High Energy
Astrophys.,} ed. J. M. Matthews, (Singapore: World Scientific), 32.

\ref Ramaty, R., Mandzhavidze, N., Kozlovsky, B., Skibo, J.
1993, \asr, {\sbf 13}, No. 9(9), 275.

\ref Ramaty, R., Skibo, J. G., and Lingenfelter, R. E. 1994, \apjs,
{\sbf 92}, 393.

\ref Ramaty, R., Schwartz, R. A., Enome, S., and Nakajima, H.
1994, \apj, {\sbf 436}, 941.

\ref Reames, D. V. 1990, \apjs, {\sbf 73}, 235.

\ref Reames, D. V., Meyer, J-P., and von Rosenvinge, T. T.
1994, \apjs, {\sbf 90}, 649.

\ref Reeves, H. 1994, {\it Revs. Modern Physics,} {\sbf 66}, 193.

\ref Reeves, H., Fowler, W. A., and Hoyle, F. 1970, {\it
Nature, Phys. Sci.}, {\sbf 226}, 727.

\ref Reeves, H., and Prantzos, N. 1995, in {\it Light Element Abundances,}
ed. Ph. Crane, in press.

\ref Remillard, R. A., McClintock, J. E., and Bailyn, C. D.
1992, \apj, {\sbf 399}, L145.

\ref Rieger, E. 1989, \sp, {\sbf 121}, 323.

\ref Riegler, G. R.  \etal 1981, \apj, {\sbf 248}, L13.

\ref Ruiz-Lapuente, P., Lichti, G. G., Lehoucq, R., Canal, R.,
and Casse, M. 1993 \apj, {\sbf 417}, 547.

\ref Ryan, J. et al. 1993, \xxiiiicrc, 3, 103.

\ref Schmelz, J. T. 1993, \apj, {\sbf 408}, 381.

\ref Schneid, E. J. 1994, Paper Presented at the 1994 AAS Winter
Meeting, Crystal City, Virginia.

\ref Schramm, D. N. 1971, \ass, {\sbf 13}, 249.

\ref Share, G. H., Kinzer, R. L., Kurfess, J. D., Forrest, D.
J., and Chupp, E. L. 1985, \apj, {\sbf 292}, L61.

\ref Shemi, A. 1991, {\it Mon. Not. Royal Astr. Soc.},
{\sbf 251}, 221.

\ref Shibazaki, N., and Ebisuzaki, T. 1988, \apj, {\sbf
327}, L9.

\ref Shigeyama, T., Nomoto, K., Tsujimoto, T., and Hashimoto, M.
1990, \apj, {\sbf 361,} L23.

\ref Skibo, J. G. 1993, {\it Diffuse Galactic Positron
Annihilation Radiation and the Underlying Continuum}, PhD
Dissertation, University of Maryland.

\ref Skibo, J. G., Dermer, C., and Ramaty, R. 1994, \apj, {\sbf
431}, L39.

\ref Skibo, J. G., and Ramaty, R. 1991, in \sacaip, 168.

\ref Skibo, J. G., Ramaty, R., and Leventhal, M. 1992, \apj,
{\sbf 397}, 135.

\ref Smith, D. M. \etal 1993, \apj, {\sbf 414,} 165.

\ref Smith, D.M., Leventhal, M., Cavallo, R., Gehrels, N.,
Tueller, J., and Fishman, G. 1995, \apj, submitted.

\ref Swinson, D. B., and Shea, M. A. 1990, {\it Geophys. Rev.
Lett.}, {\sbf 17}, 1073.

\ref Sunyaev, R.  \etal 1987,  \nature, {\sbf 330}, 227.

\ref Sunyaev, R. \etal 1991, \apj, {\sbf 383}, L49.

\ref Sunyaev, R. \etal 1992, \apj, {\sbf 389}, L75.

\ref Teegarden, B. J., Barthelmy, S. D., Gehrels, N., Tueller,
J., Leventhal, M., and MacCallum, C. J. 1989, \nature, {\sbf
339}, 122.

\ref Teegarden, B. J., Barthelmy, S. D., Gehrels, N., Tueller,
J., Leventhal, M.,  and MacCallum, C. J. 1991, in \sacaip, 116.

\ref Thielemann, F.-K., Hashimoto, M., and Nomoto, K. 1990,
\apj, {\sbf 349}, 222.

\ref Tueller, J., Barthelmy, S., Gehrels, N., Teegarden, B. J.,
Leventhal, M., and  MacCallum, C. J. 1990, \apj, {\sbf 351}, L41.

\ref van den Bergh, S. and Tammann, G. A. 1991, \araa, {\sbf
29}, 363.

\ref Vestrand, W. T., and Forrest, D. J. 1993, \apj, {\sbf 409},
L69.

\ref von Ballmoos, P., Diehl, R.,  and Schonfelder, V. 1987,
\apj, {\sbf 318}, 654.

\ref Wang, H. T., and Ramaty, R, 1974, \sp, {\sbf 36}, 129.

\ref Weaver, T. A., and Woosley, S. E. 1980a, {\it Ann. NY Acad.
Sci.}, {\sbf 336}, 335.

\ref Weaver, T. A., and Woosley, S. E. 1980b, in {\it Supernova
Spectra}, eds. R. Meyerott and G. H. Gillespie, (New York:
AIP), 15.

\ref Weaver, T. A. and Woosley, S. E. 1993, {\it Physics
Reports}, {\sbf 227}, 65.

\ref Werntz, C., Lang, F. L., and Kim, Y. E. 1990, \apjs, {\sbf
73}, 349.

\ref Winkler, C. 1994, \apjs, {\sbf 92}, 327.

\ref Woosley, S. E., Hartmann, D., Hoffman, R., D., and Haxton,
W. C. 1990, \apj, {\sbf 356}, 272.

\ref Woosley, S. E., and Pinto, P. A. 1988, in \washaip, 98.

\ref Woosley, S. E. 1991, in \sacaip, 270.

\ref Woosley, S. E., and Weaver, T. A. 1986 in {\it
Nucleosynthesis and Its Implications on Nuclear and Particle
Physics}, eds. J. Audouze and N. Mathieu, (Dordrecht: Reidel),
145.

\ref Yoshimori, M. 1990, \apjs, {\sbf 73}, 227.

\ref Yoshimori, M. \etal 1994, \apjs, {\sbf 90,} 639.

\eject

\rm

\vskip 0.4 truecm \noindent {\sbf FIGURE CAPTIONS} \vskip 0.2 truecm

\noindent Fig.~1. Calculated gamma ray deexcitation spectra;
upper panel -- a combined narrow and broad line spectrum; lower
panel -- a broad line spectrum showing the very narrow lines
from long lived radionuclei (from Ramaty 1995).
\vskip 0.2 truecm

\noindent Fig.~2. A theoretical solar flare gamma ray spectrum
showing the strongest expected nuclear deexcitation lines.
\vskip 0.2 truecm

\noindent Fig.~3. Observed solar flare gamma ray line spectrum
fitted with theoretical curve (from Murphy et al. 1991).
\vskip 0.2 truecm

\noindent Fig.~4. The observed location of the nuclear line
emission in the Orion star formation region and the spectrum of
the line emission (from Bloemen et al. 1994).
\vskip 0.2 truecm

\noindent Fig.~5. Calculated energy deposition rates in Orion
by accelerated particles of various compositions in a neutral
ambient medium with solar photospheric composition, as
functions of the accelerated particle spectral parameter $E_0$
(from Ramaty et al. 1995b).
\vskip 0.2 truecm

\noindent Fig.~6. Production ratios of $^9$Be to 3-7MeV nuclear
deexcitation photons for various accelerated particle
compositions as functions of the accelerated particle spectral
parameter $E_0$ (from Ramaty et al. 1995b).
\vskip 0.2 truecm

\noindent Fig.~7. Production ratios of $^7$Li to $^6$Li for
various accelerated particle compositions as functions of the
accelerated particle spectral parameter $E_0$. Shown are the
meteoritic ratio (Anders \& Grevesse 1989) and the two values
for clouds in the direction of $\zeta$ oph (Lemoine et al.
1995).
\vskip 0.2 truecm

\noindent Fig.~8. Production ratios of $^{26}$Al to $^9$Be for
various accelerated particle compositions as functions of the
accelerated particle spectral parameter $E_0$ (from Ramaty et
al. 1995b). The horizontal bar represents the protosolar
$^{26}$Al abundance and the probable $E_0$ range.
\vskip 0.2 truecm

\noindent Fig.~9. The 0.847 MeV line observed (Tueller et al.
1988) with a high resolution detector from Supernova 1987A
compared with calculations (Pinto and Woosley 1988) for the
mixed model 10HMM, suggesting significant asymmetry in the
ejecta.
\vskip 0.2 truecm

\noindent Fig.~10. The $^{44}$Ti line emission from the young
supernova remnant Cas A; left panel -- the location of the
line emission; right panel -- the spectrum of the
line emission (from Iyudin et al. 1994).
\vskip 0.2 truecm

\noindent Fig.~11. The sky in 1.809 MeV $^{26}$Al decay line
emission (from Diehl et al. 1995).
\vskip 0.2 truecm

\noindent Fig.~12. The 0.511 MeV positron annihilation
line observations from the region of the Galactic center
(from Leventhal et al. 1993).
\vskip 0.2 truecm

\noindent Fig.~13. Galactic longitude profiles of the 0.511 MeV line
emission (from Skibo et al. 1993).
\vskip 0.2 truecm

\noindent Fig.~14. The observed (Goldwurm et al. 1992) spectrum
from Nova Muscae during the outburst on 20 January 1991
compared with the Monte Carlo simulated (Hua and Lingenfelter
1993) spectrum of Compton scattered annihilation radiation for
an accretion disk around a black hole, superimposed on a
background power law with index $s = -2.90$ and viewed at an
observing angle of $68^{\circ}$ with respect to the axis of the
accretion disk.

\bye